\documentclass[12pt]{article}
\usepackage{format_style}
\usepackage{authblk}
\usepackage{graphicx}
\usepackage{color}
\usepackage{xcolor}
\usepackage{hyperref}
\usepackage{mathptmx} 
\usepackage{multirow}
\usepackage{booktabs}

\newcommand{\overbar}[1]{\mkern 1.5mu\overline{\mkern-1.5mu#1\mkern-1.5mu}\mkern 1.5mu}
\usepackage{amsfonts}

\newcommand{\gharana}{ghar\={a}n\={a}}

\newcommand{\Gharana}{Ghar\={a}n\={a}}

\usepackage{diagbox}
\usepackage{pdfcomment}
\usepackage{lipsum}  

\title{Structural Segmentation and Labeling of Tabla Solo Performances}

\date{\today}
\author[1]{Gowriprasad R\footnote{\href{mailto:ee19d702@smail.iitm.ac.in}{ee19d702@smail.iitm.ac.in}}}
\affil[1]{Department of Electrical Engineering, IIT Madras, Chennai 600 036, India} 
\author[1]{R Aravind\footnote{\href{mailto:aravind@ee.iitm.ac.in}{aravind@ee.iitm.ac.in}}}
\author[2]{Hema A Murthy\footnote{\href{mailto:hema@cse.iitm.ac.in}{hema@cse.iitm.ac.in}}}
\affil[2]{Department of Computer Science and Engineering, IIT Madras, Chennai 600 036, India} 

\date{\today}

\begin{document}
{\setstretch{.8}
\maketitle
\begin{abstract}

\textit{Tabla} is a North Indian percussion instrument used as an accompaniment and an exclusive instrument for solo performances. Tabla solo is intricate and elaborate, exhibiting rhythmic evolution through a sequence of homogeneous sections marked by shared rhythmic characteristics. Each section has a specific structure and name associated with it. Tabla learning and performance in the Indian subcontinent is based on stylistic schools called \emph{\gharana-s}. Several compositions by various composers from different \gharana-s are played in each section. This paper addresses the task of segmenting the tabla solo concert into musically meaningful sections. We then assign suitable section labels and recognize \gharana-s from the sections. We present a diverse collection of over 38 hours of solo tabla recordings for the task. We motivate the problem and present different challenges and facets of the tasks. Inspired by the distinct musical properties of tabla solo, we compute several rhythmic and timbral features for the segmentation task. This work explores the approach of automatically locating the significant changes in the rhythmic structure by analyzing local self-similarity in an unsupervised manner. We also explore supervised random forest and a convolutional neural network trained on hand-crafted features. Both supervised and unsupervised approaches are also tested on a set of held-out recordings. Segmentation of an audio piece into its structural components and labeling is crucial to many music information retrieval applications like repetitive structure finding, audio summarization, and fast music navigation. This work helps us obtain a comprehensive musical description of the tabla solo concert.

\noindent
\textit{\textbf{Keywords: }%
\Gharana, P\={e}\`{s}k\={a}r; K\={a}yad\={a}; Ga\d{t}; Tukd\={a}; Chakradh\={a}r.} \\ 

\end{abstract}
}

\section{Introduction}

With the availability of varied music collections on digital platforms and widespread use of personal digital devices, there is a growing interest in accessing music based on its various characteristics. Metadata supplied with the audio recordings of concert performances that are available online contain information about the musicians and the performance duration. But, the information relating to the section boundaries and other annotations are rarely provided, irrespective of whether it is vocal, accompaniment or percussion, especially in the context of Indian art music. The limited availability of editorial metadata and annotations led to the need for music information retrieval (MIR) to automatically extract characteristic properties of music from the audio recordings.



From the MIR perspective, analyzing the rhythm structure of tabla performances is vital. Automatic segmentation of concert audios offers a variety of MIR applications, including quick navigation \cite{cooper2003summarizing,peeters2003deriving}, detection of repeating structure in music, and meaningful transcription of music \cite{klapuri2001automatic}. The tasks such as audio thumbnailing \cite{bartsch2005audio}, auto-tagging, music summarization and description, similarity measurements, as well as informed and enhanced music listening, training and computational musicology, are all made easier by automatic metadata identification. Identifying metadata from audio-like, stylistic school recognition—especially in the same genre—is a difficult process for humans.

\begin{figure}[t]
    \centering
    \resizebox{\columnwidth}{!}{%
    \pdftooltip[]{
    \includegraphics[trim={0cm 0cm 0cm 0cm}, clip]{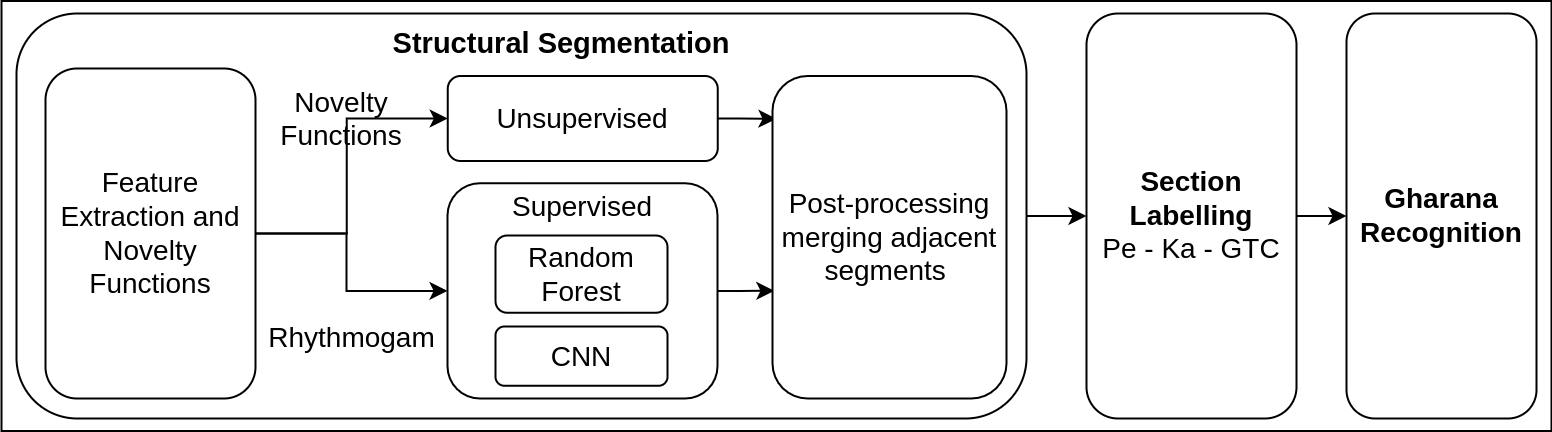}
    }{Figure shows the Overall flow depicting the three major tasks discussed in the paper.}%
    }
    \caption{Overall flow depicting the three major tasks discussed in the paper. \\ Alt-Text: Figure shows the Overall flow depicting the three major tasks discussed in the paper.}
    \label{fig:Overall_sections}
\end{figure}

This paper addresses the structural segmentation and labeling of tabla solo concerts. Tabla solo is intricate, with a variety of precomposed compositions with further elaborations based on the player's stylistic schools called \gharana-s\footnote{The word \gharana\ literally means house and implies the house of the teacher.} \cite{gowriprasad_r_gharana_ISMIR2021,bagchee1998nad}. Figure \ref{fig:Overall_sections} shows the overall tasks focused in the paper. Segmentation of a tabla solo performance involves finding the boundaries that mark the transition between segments with different rhythmic structures. Other important tasks are labeling the segments and the identification of tabla \gharana-s on the segmented audio clips. The paper addresses three tasks: 1. Concert segmentation, 2. Segment classification and labeling, 3. \Gharana\ recognition and labeling of each segment. We perform genre/culture-specific analyses since tabla solo performances are distinct from other genres. The paper explores techniques used in other well-researched genres of music, with culture-specific changes for both supervised and unsupervised structural segmentation and labeling methods.



The rest of the paper is structured as follows. In the following part of the current section, we discuss an overview of tabla solo performances and tabla \gharana-s, followed by a literature review. The challenging factors influencing each task are categorically mentioned motivating the experiment. Section \ref{sec2:dataset_Description} describes the dataset collected as well as the ground truth annotation procedure. Section \ref{sec3:Audio_and_Features} discusses the pre-processing and feature extraction steps, inspired by the characteristics of the music tradition. The proposed segmentation methods, and models are explained in Section \ref{Sec4:Segmentation_methods}. Section labeling and \gharana\ recognition tasks are described in Section \ref{sec5:Section_Classification_and_Labeling}. Multiple experiments addressing various facets of the task are described. The experimental results are analyzed and discussed in Section \ref{sec6:Experiments_and_Results}. Finally, Section \ref{sec7:Conclusion} summarizes the work mentioning the key takeaways.




\subsection{Structure of Tabla Solo Concert and \Gharana-s}
\label{subsec:concert_structure}

Tabla solo is entirely independent of the Hindustani vocal or other instrumental concerts. Tabla is the main instrument, and the instruments like Sarangi, Harmonium, or sometimes Violin are used for \emph{Lehra} accompaniment. The purpose of lehra is to indicate the reference metric tempo \cite{bagchee1998nad} while providing a background melody. Tabla solo is intricate, with a variety of pre-composed forms within the rhythmic framework called \emph{t\={a}l}. Further elaborations are based on the player's school of practice called \gharana.



Tabla solo consists of different compositions such as \emph{\d{t}h\={e}k\={a}, u\d{t}h\={a}n, p\={e}\`{s}k\={a}r, k\={a}yad\={a}, r\={e}l\={a}, rau, ga\d{t}, para\d{n}, tukd\={a}, chakradh\={a}r} \cite{pradhan2011tabla}. Each composition has different functional and aesthetic roles in a solo performance. The structural framework of a tabla solo is generally as follows: The \emph{lehra} accompaniment plays an al\={a}p initially for a few minutes. Then the tabla player starts with the initial u\d{t}h\={a}n or directly p\={e}\`{s}k\={a}r. P\={e}\`{s}k\={a}r rendition can go from five minutes to more than 15-20 mins depending on the overall concert duration. Next to p\={e}\`{s}k\={a}r, various k\={a}yad\={a} compositions from different \gharana-s are played one after the other. Further, the compositions such as r\={e}l\={a}, rau are presented. Later on, the metric tempo increases and the fixed compositions such as ga\d{t}-s, stuti para\d{n}-s, tukd\={a} and mukh\d{d}\={a}-s are played. In the end, the chakradh\={a}r-s are played, which are essentially long-term mukh\d{d}\={a}-s. Therefore, the sequence of these pieces would broadly be as follows: \emph{p\={e}\`{s}k\={a}r/u\d{t}h\={a}n → k\={a}ydha → r\={e}l\={a} → ga\d{t} → tukd\={a}/chakradh\={a}r}



It is necessary to point out the role that is played by the \d{t}h\={e}k\={a}\footnote{\d{t}h\={e}k\={a} is the basic stroke pattern associated with a particular rhythm cycle "t\={a}l"} in the context of solo tabla recitals. The basic \d{t}h\={e}k\={a} is usually played in between adjacent compositions, marking the start and end of the item \cite{pradhan2011tabla}. This indicates the basic tempo (the bar\={a}bar laya) on which the repertoire and subsequent improvisation are built. This is the metric tempo as indicated and maintained by \emph{lehra} accompaniment. After establishing the metric tempo by playing \d{t}h\={e}k\={a}, the player can perform further elaborations as the performance progresses. The \emph{thek\={a}} thus provides the standard for improvisations \cite{bagchee1998nad}. Another aspect is that the choice of metric tempo is highly flexible, and the artiste decides it on stage. The metric tempo changes across the concerts, as well as for different compositions within the concert. The tempo is measured in beats per minute (bpm).

\subsubsection{Major sections in tabla solo}
\label{subsec:structure_tabla_solo}
Based on the item/compositions' structure and their position in the recital, one can find three major sections in the tabla solo concert. They are p\={e}\`{s}k\={a}r (Pe), k\={a}yad\={a} (Ka), and ga\d{t}--tukd\={a}--chakradh\={a}r (GTC) sections, each of these sections have a specific meaning as indicated below.

Pe: The word p\={e}\`{s}k\={a}r means "to present" and is played at the beginning of the concert. Through the p\={e}\`{s}k\={a}r, the performer establishes the basic structure of the t\={a}l that is being presented. It sets the mood of the t\={a}l and of the performance itself. Many tabla players liken it to the al\={a}p in vocal and instrumental recitals. P\={e}\'{s}k\={a}r is an extempore rendition played at a slower metric tempo (40-45bpm). It is usually the single longest section in the concert.


Ka: The k\={a}yad\={a}, meaning "rule," follows a theme and consists of a pattern of variations, where the variations use syllables from the theme. Thus, the structure is restrictive and demands that the tabla player works within bounds. K\={a}ydha-s are usually played in a medium metric tempo (55-65bpm). Each k\={a}yad\={a} ends with a tih\={a}yi which is essentially a small phrase that is repeated thrice. Each k\={a}yad\={a} rendition usually lasts for two to five minutes.

GTC: The ga\d{t} is a short and pithy composition, which generally employs loud and heavy strokes influenced by the Pakh\={a}waj\footnote{A double barrel drum used as rhythm accompaniment in Hindustani music.} vocabulary. The tukd\={a}, and chakradh\={a}r are non-extendable pre-conceived compositions played towards the end of a solo recital. The chakradh\={a}r structure involves a short passage followed by a tih\={a}yi, but the entire unit has to be repeated thrice for it to resolve on the first beat of the approaching cycle called as ''sam''. The GTC are examples of fixed compositions played at a fast metric tempo (100-130bpm).


\begin{figure}[t]
    \centering
    \resizebox{\columnwidth}{!}{%
    \pdftooltip[]{
    \includegraphics[trim={0cm 0cm 0cm 0cm}, clip]{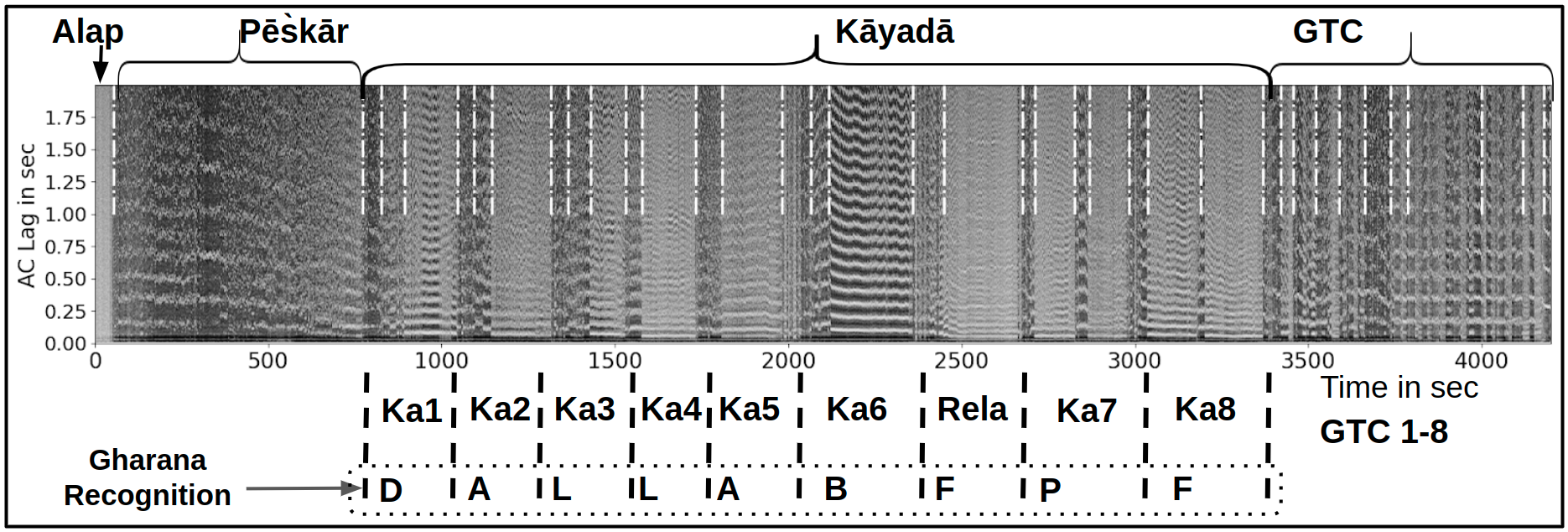}
    }{Figure shows the overall objective illustration by depicting the three tasks}%
    }
    \caption{Objective Illustration. Task 1: Getting segment boundaries (vertical dotted lines), Task 2: Section labeling (p\={e}\`{s}k\={a}r, k\={a}yad\={a}, and GTC), Task 3: Gharana recognition on the segment audios. The feature illustrating the structure of tabla solo is called the rhythmogram which is described in Section \ref{subsec:Rhythmogram}. \\ Alt-Text: Figure shows the overall objective illustration by depicting the three tasks}
    \label{fig:Overall_Objective}
\end{figure}

\subsubsection{Tabla \Gharana-s}

Teacher-student lineage gave rise to six different \gharana-s namely Delhi (D), Ajrada (A), Lucknow (L), Banaras (B), Farukhabad (F), and Punjab (P). Each \gharana\ is characterized by its unique style of playing \cite{pradhan2011tabla}, strokes, improvisations, and pre-composed patterns. Each stylistic school also has differences in elaboration and dynamics of tabla phrases and specific compositional patterns (signature patterns). Compositions are the heart of tabla's repertoire. These compositions are usually from different \gharana-s. Thus we consider the aspect of repertoire, in so far as it has a bearing on the \gharana\ distinctions \cite{bagchee1998nad}. For more details of tabla solo and \gharana-s, refer to \cite{bagchee1998nad,pradhan2011tabla,Tabla_Vidwat,saxena2006art,gottlieb1993solo}. Figure \ref{fig:Overall_Objective} illustrates the overall objective of the paper. It depicts the segmentation boundaries, section labels along with \gharana\ annotations.

\subsection{Related works}

Musical attributes like timbre, melody, rhythm, or harmony represent the musical structure differently. Different acoustic and temporal parameters capture these musical characteristics and are used to determine the structural boundaries in music. In Western music, many researchers have used timbre features as harmony-based features for music segmentation into intro-chorus-verse-outro sections \cite{dannenberg2008music,paulus2010state}. \cite{grosche2010cyclic} used Low dimensional tempo features for segmentation tasks in Western classical music. Several inherent challenges in the structural segmentation of classical sonata music were reviewed by \cite{allegraud2019learning}. The authors used melodic, harmonic, and rhythmic features that characterized the evolving structure and formulated them as a sequence of recurring states.

There are many different approaches to music segmentation. A few approaches exploit the homogeneity of the sections in some musical aspects. A few novelty-based approaches are based on detecting a sudden change in the musical properties, and the recursion-based approaches detect the repeating structure in the music. Different unsupervised approaches were reviewed by grouping similar sections using structural analysis \cite{paulus2010state}. \cite{foote2000automatic} proposed a method that uses a self-distance matrix (SDM) to determine the boundary between contrasting musical characteristics. \cite{turnbull2007supervised} characterize the changes in musical features like melody, harmony, timbre, and rhythm in Pop and Rock music to train the boosted decision stump (BDS) supervised classifier for predicting boundary frames. Lately, \cite{ullrich2014boundary} explored the mel-scaled spectrograms in training the convolutional neural networks (CNN) for the structural segmentation on the SALAMI dataset that spans a large variety of genres \cite{smith2011design}.

In the context of Indian music, different approaches were explored for the segmentation task, addressing both the Hindustani and Carnatic music traditions. \cite{padi2018segmentation} identified the applause instances between adjacent compositions in live recordings of the Carnatic music as a cue for segmentation. In Indian art music owing to its improvisational nature, it is common to hear applause intra item. Nevertheless, melodies are defined by ragas, and the same melody will not be used in adjacent items. This property is exploited in this work to segment continuous recordings of concerts into items. The authors used tonic normalized timbre features (cent filterbank cepstral coefficients - CFCC) and the pitch histograms in merging the adjacent sections. \cite{pv2016segmentation} addresses the segmentation of carnatic music items into alapana and kriti using GMMs and energy features. \cite{sankaran2015automatic} also used CFCC feature templates to segment the compositions in Carnatic music automatically. \cite{verma2015structural} explored the structural segmentation of Hindustani instrumental and Dhrupad concert recordings \cite{rohit2020structure}. In the case of instrumental concerts, Vinutha et al. \cite{vinutha2016structural} considered the segmentation of sitar and sarod concerts using reliable tempo detection \cite{vinutha2016reliable}. In the case of Dhrupad vocal concerts, \cite{rao2020structural} performed the segmentation of alap, jod, and jhala in dhrupad vocal concerts. They have also segmented Dhrupad bandish using the tempo \cite{ma2020structural}. \cite{ranjani2013hierarchical} considered the aspects of rhythmicity and percussiveness, strongly signaled by the accompanying percussion instrument's onsets and low frequency content to classify the concert sections in Carnatic music. \cite{thoshkahna2015novel} exploited the salience of the estimated tempo to distinguish sections with ambiguous tempo (alapana) from the later concerts sections with clear rhythmic properties in Carnatic music concerts.

Indian percussion research has mainly focused on stroke transcription \cite{gillet2003automatic,chordia2005segmentation} and sequence modeling \cite{chordia2011predictive,chordia2010multiple} of tabla strokes. The stroke transcription task was developed using hidden Markov models \cite{samudravijaya2004computer}, followed by percussion pattern identification~\cite{gupta2015discovery}. Recently, \cite{rohit_ismir_2021} performed four-way classification of tabla strokes using the models adapted from automatic drum transcription. In case of mridangam, stroke transcription can be found in \cite{kuriakose2015akshara,anantapadmanabhan2013modal}. \cite{srinivasamurthy2014search} explored the rhythmic analysis of Indian and Turkish music where the authors addressed the meter tracking and beat tracking tasks. 

The works on music style and classification are as follows. Melodic contours were used to classify vocal style \cite{vidwans2012classification}, and the melodic features for classifying cultural music \cite{vidwans2020classifying}. In case of percussion, the task of mridangam artiste identification from mridangam tani-avartanam audio was attempted \cite{gogineni2018mridangam}. Recently, tabla \gharana\ recognition from the solo tabla performance audios was addressed in \cite{gowriprasad_r_gharana_ISMIR2021,gowriprasad_r_gharana_NCC2022}.

\subsection{Motivation and challenges}

We motivate the task of segmentation and labeling by getting insights into the factors influencing each sub-tasks. The factors include both the supporting and challenging aspects. We consulted four tabla maestros to get expert advice on the factors influencing the task. Based on the discussion and the artists' common opinions, we formulate the challenging factors for each of the sub-tasks.

\textit{Structural Segmentation:} Tabla solo is a structured rendition of numerous compositions played one after the other. The number of compositions played and the duration of each composition is also not fixed. Thus the number of segments in each performance varies. There exist some instances of spoken recitation of the compositions before being played. Each composition has its own rhythmic structure and is presented at multiple speeds. This is reflected in the boundary within a single composition itself due to sudden changes in tempo or the rhythmic structure. The rendition also has pauses, which may be part of the composition itself or due to the artiste's presentation style. Thus the tabla solo segmentation task poses new challenges to the existing audio segmentation methods. Listening to the entire audio carefully to mark the segment boundaries is time-consuming. This also motivates us to build systems for automatic segmentation and annotations.

\textit{Section labeling and \Gharana\ recognition:} Artistes nowadays would have learned from several teachers from different \gharana-s. Different \gharana\ styles will also influence their playing style \cite{bagchee1998nad}. Thus the structure of the tabla solo and its components have also evolved. There is no specific rule governing a certain composition's position in the concert. As mentioned in Section \ref{subsec:structure_tabla_solo}, the tabla solo concerts consist of three kinds of renditions: mostly extempore and extempore nuances upon a theme and fixed compositions. It is hard to label the rendition just by stroke patterns and structures. 


Tabla is a pitched harmonic percussive instrument tuned to a specific tonic in a concert  \cite{saxena2006art,anantapadmanabhan2014tonic}. As the tonic varies, the properties of the sound, like harmonics, timbre, tone, etc., also vary. Thus the feature vectors representing the same stroke with different tonic will also change, challenging the system performance. In an ideal scenario, the metadata is independent of tonic variability. \cite{gottlieb1993solo} mentions three factors for comparing the similarities and differences in playing: (1) Sound production, that is, quality and the technique used, (2) Repertoires, and (3) Rhythmic practices. The technique and the rhythmic practices differ from artiste to artiste. Compositions bearing the \gharana\ distinctions are based on the repertoires of each \gharana-s \cite{gowriprasad_r_gharana_ISMIR2021}. The challenges involved in each task motivate addressing the unexplored problem of segmentation and labeling tabla solo recordings. 

\section{Dataset Description}
\label{sec2:dataset_Description}
\vspace{-0.05cm}



Since there was no tabla solo dataset available, we collected tabla solo recordings from commercial audio CDs, live recordings from the artistes' archives, and online sources. This corpus consists of tabla solos played in 5 different t\={a}l-s by 25 artistes. All the artistes are senior exponents from different tabla \gharana-s with at least 20 years of tabla playing experience. A concert's duration varies from 15 minutes to 80 minutes. Concert audios were annotated for the major structural boundaries. The overall dataset comprises 55 solo tabla audios with a total duration of around 38 hours, having more than 2000 segment boundaries.

\subsection{Labeling the raw data}

The rhythm structure and tempo are the most distinctive properties of a segment within a concert, and the relative change in tempo and rhythmic structures serves as cues for segment boundary detection. A few cycles of \d{t}h\={e}k\={a} are played in between adjacent compositions, marking the start and end of a composition \cite{pradhan2011tabla}. Each compositional theme is played initially at a speed of or double the speed of the original tempo (metric tempo) for one or two cycles and is then played at four times the original tempo \cite{pradhan2011tabla}. A step increase in surface tempo is observed at the initial stages of the compositions. The sharp decrease in the stroke density (indicator of surface tempo) is usually observed at the end of each composition to \d{t}h\={e}k\={a}. This is depicted in the Figure \ref{fig:Rhythmogram_1} and \ref{fig:Rhythmogram_2}. 


Professional performers were employed to listen and extract the k\={a}yad\={a}, ga\d{t}, and chakradh\={a}r sections from the audio by marking the start and endpoints. The ground truth annotations relied on various culture-specific cues mentioned above. The aspect of tempo change is recognized immediately by the listeners; the section boundaries at the end of the composition transiting from the higher speed to the lower speed \d{t}h\={e}k\={a} are marked consistently. This ensured the reliability of the labeling. Then four tabla maestros from different \gharana-s were requested to listen to these audio segments and give the ground truth labels for the section name as well as the \gharana. Manual annotations are subjective due to the multiple musical cues. We discussed with four tabla maestros and followed a consensus-based approach in deciding the boundary annotations. It is to be mentioned that the performers had to listen to the entire audio carefully to mark the segment boundaries, which was time-consuming. This also motivates us to build systems for automatic segmentation and annotations.

Structural segmentation in the context of tabla solo involves the detection of start and end instances of compositions. The task is not addressed at the metrical time scale, or stroke level but at a larger time scale. Thus the tolerance duration is not in milliseconds as in the case of stroke onset detection \cite{gowriprasad2020onset,bello2005tutorial} but in a larger time scale in "seconds". For completeness, the segmented portion is expected to contain the full composition. The start marking needs to be before the composition starts, and the end markings need to be after the composition ends. Marking the segment boundaries precisely on the first beat of the t\={a}l cycle where the compositions start and end will affect the completeness for the listeners. Usually, in practice, the artists will play a few cycles of th\={e}ka before the start of any composition (k\={a}ydha, GTC). This establishes the t\={a}l structure and the metric tempo for the composition to be played. Thus the smaller segments in between the compositions where th\={e}ka is played for a few cycles are included with the following segment.



\subsection{Statistical analysis of the annotated data}
\label{sec:statistical_analysis}
\begin{figure}[t]
    \centering
    \resizebox{\columnwidth}{!}{%
    \pdftooltip[]{
    \includegraphics[trim={4cm 0.67cm 1.6cm 1cm}, clip]{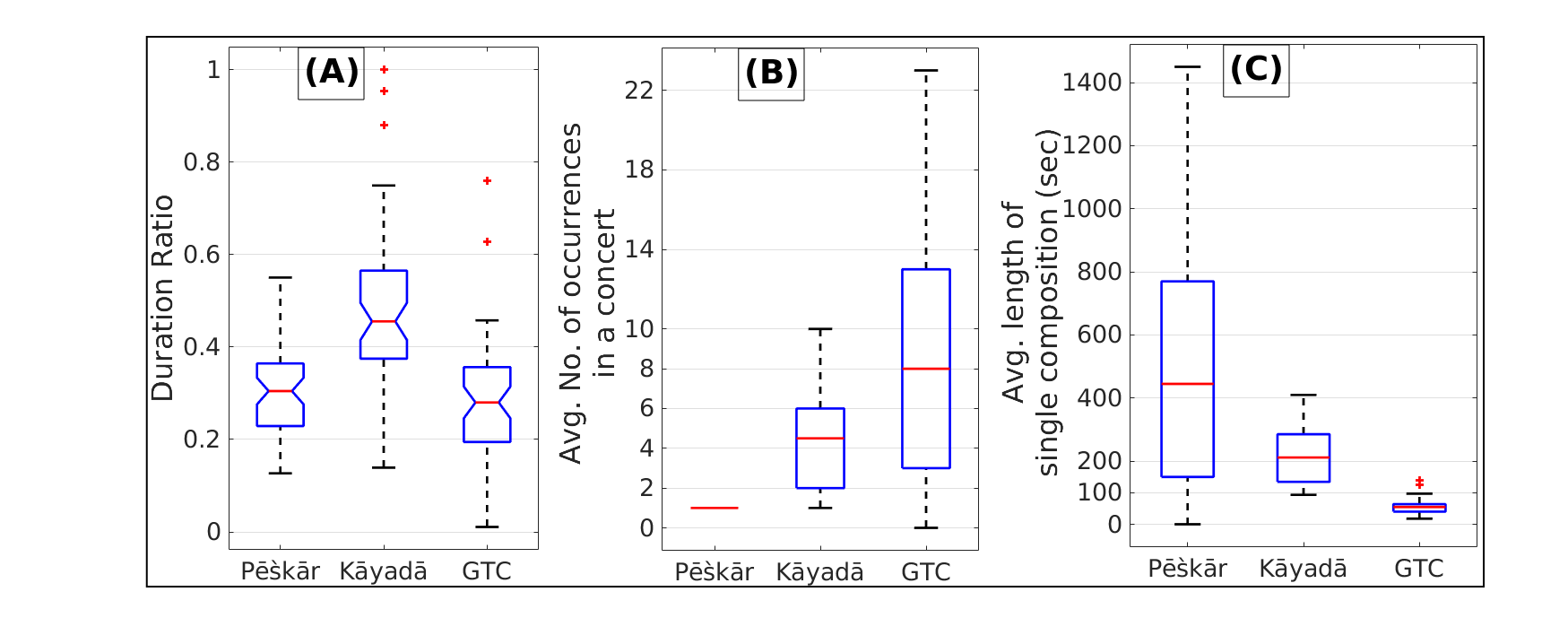}
    }{Figure shows boxplots plotted on three different parameters of the dataset}%
    }
    \caption{Statistical analysis of the annotated data. \\ Alt-Text: Figure shows boxplots plotted on three different parameters of the dataset}
    \label{fig:mean_variance_data}
\end{figure}

Plots in Figure \ref{fig:mean_variance_data} show the various statistics of three sections of tabla solo computed from the entire dataset. The first plot ($A$) showcases the duration ratio of each section in the concert. The average duration ratio Pe-Ka-GTC is $0.3,0.48,0.28$, respectively. The second plot ($B$) showcases the average number of individual compositions in each section in a concert. The p\={e}\`{s}k\={a}r is one single section which is extempore. The number of k\={a}yad\={a}-s, ga\d{t}, tukd\={a}, and chakradh\={a}r compositions varies across the concert. The mean average number of k\={a}ydha-s and GTCs are $4-5$ and $8$, respectively. One can observe a large variance in the number of individual compositions in the GTC section and a relatively smaller variance in the number of k\={a}ydha-s. This essentially depends on which \gharana\ the artiste belongs. Some \gharana\ players give more emphasis on k\={a}ydha-s and some on GTC. The third plot ($C$) shows the average section length in each composition. Since p\={e}\`{s}k\={a}r is one single composition and spans nearly 30\% of the entire concert, the mean average length of p\={e}\`{s}k\={a}r is large compared to the other two sections. The mean average lengths of individual k\={a}ydha-s, and GTCs are around 210s and 60s, respectively. The larger variance in the length of p\={e}\`{s}k\={a}r in plot ($C$) and the number of k\={a}ydha-s, and GTCs in plot ($B$) also depends on the overall duration of the concert. The duration of the concert is not fixed. As mentioned earlier, in the dataset curated, the concert duration varies from fifteen and eighty minutes respectively





\section{Audio processing and Features}
\label{sec3:Audio_and_Features}

The raw concert audios have to be pre-processed for further analysis. Since each concert is unique in the choice of metric tempo, tabla tonic, and compositional structure, the feature representation need to handle concert specific characteristics as well as scale to multiple concerts. We addressed the tasks by computing relevant features considering the culture-specific musicological perspectives.

\subsection{Audio Pre-Processing and Onset Detection}

From Section \ref{subsec:concert_structure}, we know that there may exist an initial introductory speech in the case of a live concert. Mel-frequency cepstral coefficients (MFCC) are used to localize the vocal parts in the entire recording. We check for the presence of vocals only in the initial section and discard them if any. The short vocal passages in the middle of the recordings are retained as they are the instances of spoken recitation which are common in tabla solo concerts.

Initially, the raw audio is pre-processed by computing the Hilbert envelop of the linear prediction residual on the raw audio as described in \cite{gowriprasad2020onset}. Then the onset detection function (ODF) is computed using spectral flux method \cite{dixon2006simple}. The quality of the onset detection function is crucial for the analysis of tempo and rhythm. The performance of onset detection was evaluated independently on the onset annotated tabla solo audio datasets from \cite{gowriprasad2020onset,rohitacoustic,gupta2015discovery}. The best performing peak-picking threshold gave an F-score of 0.96. The computed onset locations with this threshold are considered for further analysis.



\subsection{Feature computation}
The change in the rhythm structure or the tempo is a prominent indicator of the transition between sections. We present rhythm-based and other related features motivated by observations on the tabla solo concert audios for our segmentation task. The the following feature vectors are computed for every frame, where the frame rate is two-frames/second.

\subsubsection{Rhythmogram}\label{subsec:Rhythmogram}

Rhythm is a fundamental dimension of music, and each rhythm pattern has its own time signature. In the case of percussion instruments like tabla, rhythm pattern refers to the aspects of stroke patterns. Changes in rhythm can come about from a difference in the manifestation of t\={a}l, tempo, or even a change in the surface stroke pattern. A rhythm representation of a tabla solo audio can be obtained by considering the stroke ODF (sampled at 10 ms) over a suitably long window. The inherent periodicity in the ODF is captured by the auto-correlation function (ACF). The ACF obtains the correlation between an ODF and its time-shifted version (where the shift is measured in seconds and termed the ‘lag’). The local tempo and rhythmic structure are obtained by periodicity estimation to a fixed length window over the ODF centered at the time instant of interest. The periodicity analysis using the auto-correlation function on the ODF gives a nice rhythmic representation of the audio called rhythmogram \cite{jensen2006multiple}. Rhythmogram is essentially short-time ACF strength versus time and lag axes. The ACF on the ODF is computed frame-wise with a frame length of 4 seconds and a frameshift of 0.5 seconds up to a lag of 2 seconds. Figure \ref{fig:Overall_Objective} shows the rhythmogram representation of a full concert spanning 70 minutes as it varies in time.

In the al\={a}p section, we can see that there is no periodic structure. In the remaining sections, the horizontal striations along the lag axis (y-axis) suggest the stroke periodicity. The decreasing distance between striations implies a faster rate of stroke onsets, i.e., faster tempo. The rhythmogram clearly depicts the boundaries between the segments. Thus the rhythmogram can be used as a potential feature for segmentation.

\subsubsection{Tempo — Average stroke density (ASD)}
\label{subsec:tempo_ASD}
Different compositions in tabla solos are performed at different speeds though the underlying metric tempo may remain constant. Here the metric tempo refers to the tempo of underlying lehra. The ODF captures the stroke onsets and not the \emph{lehra} variations. It is not straightforward to get the metric tempo information from the stroke onset information. The different tabla compositions have different stroke patterns and often have unequal stroke distribution across the m\={a}tras, and getting the surface tempo is not possible unless we know the stroke pattern and the speed. The tempo estimation by using the product of ACF-DFT or by counting the number of peaks along the lag axis of ACF is often prone to tempo octave errors due to uneven distribution of strokes. Thus, we compute the average stroke density (ASD) instead of computing the surface tempo. 

\begin{figure}[t]
    \centering
    \resizebox{\columnwidth}{!}{%
    \pdftooltip[]{
    \includegraphics[trim={6.1cm 0cm 5cm 0.7cm}, clip]{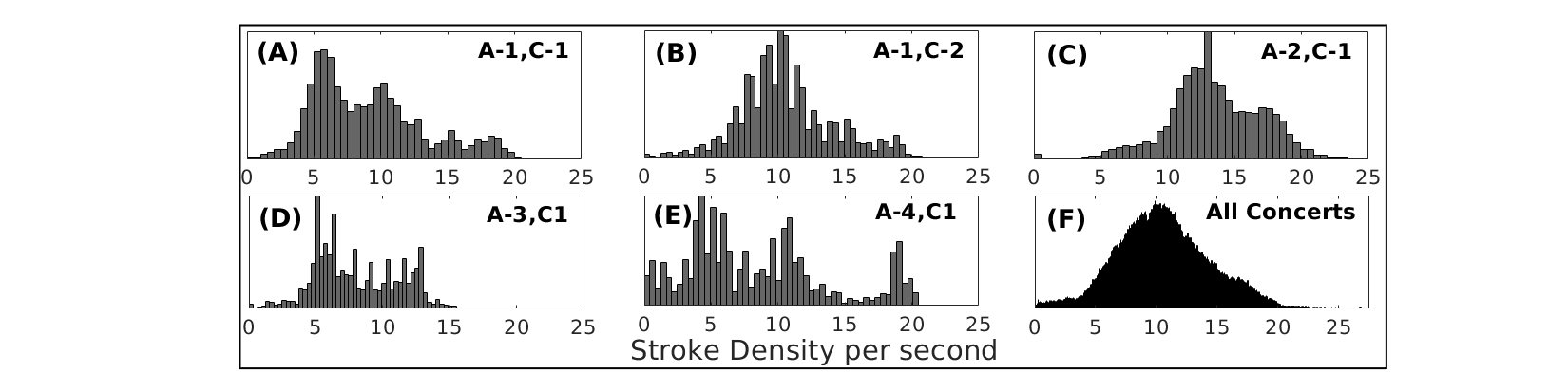}
    }{Figure shows Average stroke density histograms for different individual concerts and for all the concerts combined.}%
    }
    \caption{Histogram of stroke density per second for different concerts (A - E), and over the entire dataset (F). Mean=10.62, Median=10.38, and Standard Deviation=3.9. \\ Alt-Text: Figure shows Average stroke density histograms for different individual concerts and for all the concerts combined.}
    \label{fig:Train_loss_Hist}
    \vspace{-0.5cm}
\end{figure}

The ASD is computed on a frame-by-frame basis by counting the number of onsets in each frame. Frame size and frameshift are the same as ACF computation 4 and 0.5 seconds. This gives the ASD per 4 seconds for every 0.5 seconds. The values in each frame are divided by four to get the ASD per second. The ASD is robust to tempo octave errors and is a nice representative of surface tempo in the case of percussive audio. Figure \ref{fig:Train_loss_Hist} shows the histogram depicting the statistical estimate of stroke density per second for different concerts (A to E) and over the entire dataset (F). The subplots (A) and (B) are from the same artist (A-1) for two different concerts (C-1, C-2). The subplots (C) to (E) are from three different artistes (A-2, A-3, A-4). We can observe that the ASD is not consistent across the concerts, even for the same artiste. This is essentially due to the fact that ASD inherently depends on the compositions played. The mean and median of stroke density per second, as obtained in the entire dataset (F), are found to be $10.62$ and $10.38$, respectively. The spread of ASD also depicts the vast diversity of the dataset. We can observe the evolution of ASD over time in Figures \ref{fig:Rhythmogram_1}, and \ref{fig:Rhythmogram_2}.

%



\subsubsection{Rhythm Posteriors}

Transforming the feature vectors into a vector of class-conditional probabilities, the posteriors are shown to improve the homogeneity within the segment \cite{verma2015structural}. Since segmentation is based on the change in rhythmic structure and tempo, modeling the rhythmogram vectors to get the class-conditional posteriors helps model each section differently from adjacent sections. The intuition behind using rhythmogram features is as follows. The high dimensional rhythmogram clearly represents the homogeneity within the section and the changes between the adjacent sections. The peaks along the lag axis of the rhythmogram depicts the periodicity of the surface rhythm indicate the surface tempo \cite{vinutha2016structural}. This allows us to use Gaussian mixture models (GMM) to model the section's homogeneity and tempo.


Each feature vector of the rhythmogram $\overbar{X}_i$ is transformed to a vector $\overbar{q}_{i}$. The dimension of $\overbar{q}_{i}$ is the number of different possible speeds. The feature vectors are clustered in an unsupervised manner using a Gaussian mixture model with $k$ Gaussians representing different ACF speeds. The posterior is computed from each of the $k$ Gaussians and stacked to form a vector.


\begin{equation}
    \overbar{q}_{i}=(P\left({C_1}\mid{\overbar{X}_i}\right),P\left({C_2}\mid{\overbar{X}_i}\right),..,P\left({C_k}\mid{\overbar{X}_i}\right))
    \label{eq:posterior}
\end{equation}

In Equation \ref{eq:posterior}, the $k^{th}$ dimension of $\overbar{q}_{i}$ represents the posterior probability P, given the frame vector $\overbar{X}_i$, of the $k^{th}$ Gaussian component. The GMM is trained with maximum likelihood across all the rhythmogram frames in a given concert. Posteriors for each concert feature vector are computed from the GMMs trained individually for each concert. The noisy features affecting the homogeneity of the segment are expected to have low probability values in the mapped posterior vector. 


The major empirical parameter in modeling the GMM is the number of Gaussians. We did two tests to decide the number of Gaussians. At first, K-means clustering is performed on the feature vectors with a different number of clusters $M$ to initialize the means of GMMs. The task is to choose an optimal value of $M$ such that the total intra-cluster variation (known as a total within-cluster variation)  \(Var = \sum\limits_{m=1}^M \sum\limits_{\overbar{x}_i \in C_m} (\overbar{x}_i - \overbar{\mu}_m)^2\) is minimized. $\overbar{x}_is$ are the data points from the cluster $C_m$, and $\mu_m$ is the mean of the cluster $C_m$. The total within-cluster variation is computed for different values of $M$ and plotted. The number of Gaussians $(k)$ are chosen as the value of $M$ around the elbow point on the curve.


We quantized the ASD starting from 3 levels to 7 levels (value of $k$ around the elbow point) of quantization. It was found that 5 to 6 quantization levels were sufficient to quantize ASD with a minimum difference of 5 strokes per second. Thus we fixed the number of Gaussians to be 5. The GMM is fit only on the ACF vectors from a particular concert. Therefore, a single Gaussian is used to model one specific speed. One Gaussian for al\={a}p where ASD is nearly zero. The 2nd Gaussian corresponds to the first speed w.r.t metric tempo. 3rd Gaussian corresponds to the second speed, 4th one for fourth speed rendition. 5th one for any other speed. Thus each Gaussian depicts a different speed.

As mentioned in Section \ref{subsec:structure_tabla_solo}, the metric tempo varies across the concerts. From Section \ref{subsec:tempo_ASD}, and Figure \ref{fig:Train_loss_Hist}, we can also observe the variability of the ASD (indicator of surface tempo) across concerts. Considering these variables, we built the GMMs and computed the rhythm posterior features on each concerts separately.


\subsection{Timbre Features}

The stroke density and the stroke combinations change from composition to composition. As the energy in a frame is proportional to the number of strokes in the frame, the short time energy evolves as the ASD changes. Different strokes have a different timbre, so as the combination of strokes. As the stroke combinations change, the timbre feature characteristics also change. Thus we explore the use of MFCCs and short-time energy (STE) as the timbral features. 



MFCC + derivatives (MFCC\_$\Delta$) + double derivatives (MFCC\_$\Delta\Delta$) features and the short term energy (STE) features are extracted with a window size of 25 ms and a frameshift of 5 ms. Each MFCC\_$\Delta\Delta$ feature vector is of dimension $\mathrm{d=57}$. It is the sum of MFCC (19)+ $\Delta$+ $\Delta\Delta$ (d=19+19+19=57). The features vectors are then averaged over a 2s window with 0.5 s frameshift. Averaging over longer window filters out the noisy fluctuations at the larger timescale. We fixed the frameshift consistent with that of the rhythmogram.


\section{Structural Segmentation of Tabla Solo}\label{Sec4:Segmentation_methods}


In the context of tabla solo, structural segmentation is the task of detecting the start and end instances of compositions in a rendition that mark segment boundaries as indicated by several musical attributes. In this work, we consider music segmentation based on locating boundaries by detecting changes in the local rhythmic structure of segments at the highest timescale, that is, concert segments marked by prominent changes in rhythmic structure. The segmentation task can also be seen as a boundary detection task. A frame is classified as either a boundary or non-boundary frame based on whether a transition between segments occurs over the frame duration.

We propose two approaches for the segmentation task. The first is an unsupervised signal processing-based approach, and the second is based on supervised machine learning. The unsupervised framework involves the similarity measures and kernel correlation, and the supervised approach involves a random forest and a deep-learning CNN classifier that classifies the boundary and non-boundary frames.


\begin{figure}[t]
    \centering
    \resizebox{\columnwidth}{!}{%
    \pdftooltip[]{
    \includegraphics[trim={2cm 0cm 3cm 0.5cm}, clip]{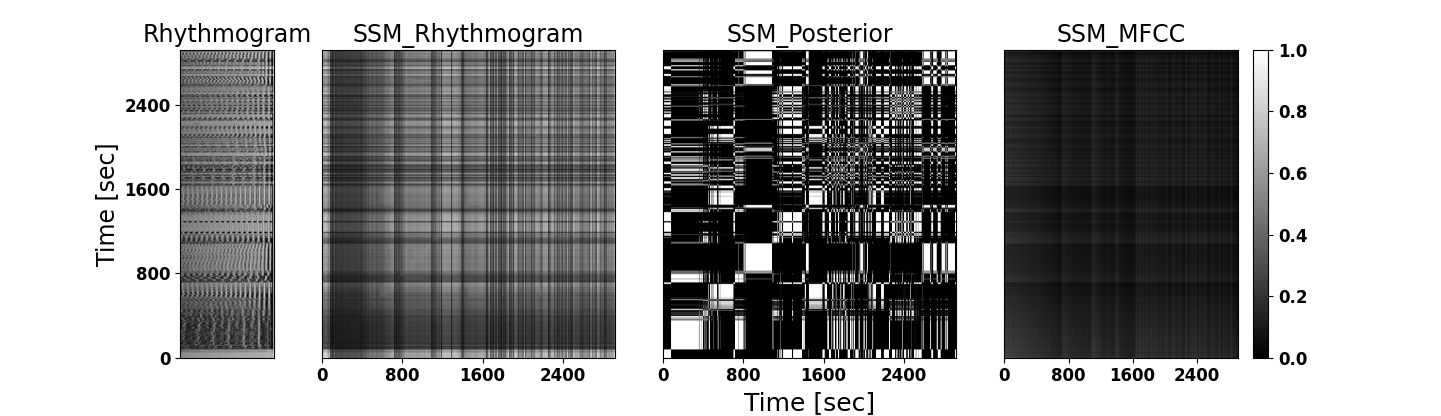}
    }{Figure shows the Self-Similarity Martices obtained from Rhythmogram, Rhythm Posteriors and MFCC features.}%
    }
    \caption{Self-Similarity Martices obtained from different features. \\ Alt-Text: Figure shows the Self-Similarity Martices obtained from Rhythmogram, Rhythm Posteriors and MFCC features.}
   \label{fig:SSMs}
\end{figure}

\subsection{Unsupervised signal processing approach}
\label{subsec:unsupervised_approach}
Given different sets of feature vectors, the self-similarity matrices (SSM) are computed on each of them using $L_2$ distance measure \cite{paulus2010state}. The homogeneous segments of length $L$ frames possibly appear as $(L \times L)$ blocks having lower distance scores. The SSM obtained on the rhythmogram, posteriors, and MFCCs are displayed in Figure \ref{fig:SSMs}. The section change points with high contrast in SSM are captured by convolving a checker-board kernel of the desired size of interest along the diagonal of SSM \cite{foote2000automatic}. The minimum segment duration is at least 10-15 seconds, as known from the annotation process (see Section \ref{sec:statistical_analysis}). The frame resolution of the feature vectors and SSM is 0.5 s. Thus, we examine $(50 \times 50)$ kernel size corresponding to $(25 s \times 25 s)$.

The 1D output obtained from the convolution is called a novelty function (NF). The peaks of the NF indicate the section boundary instances having high contrast in SSM. The aim is to get an NF whose peaks indicate the desired segment boundaries. We compute six different novelty functions from different features, as mentioned in Table \ref{tab:Novelty_functions}.


\begin{table}
\centering
\caption{Six different novelty functions from different features.}
\begin{tabular}{lc} 
\toprule
\multicolumn{1}{c}{\textbf{Features}}           & \textbf{Label}  \\ 
\hline
Hilbert envelope of the derivative of ASD       & ASD-D           \\
Hilbert envelope of the derivative of STE       & STE-D           \\
Hilbert envelope of the rhythmogram flux        & NF-RF           \\
NF from SSM on rhythmogram                      & NF-R            \\
NF from SSM on posteriors                       & NP-P            \\
NF from SSM on MFCC                             & NF-M            \\
\bottomrule
\end{tabular}
\label{tab:Novelty_functions}
\end{table}

\begin{figure}[t]
    \centering
    \resizebox{0.77\columnwidth}{!}{%
    \pdftooltip[]{
    \includegraphics[trim={0cm 0cm 0cm 0cm}, clip]{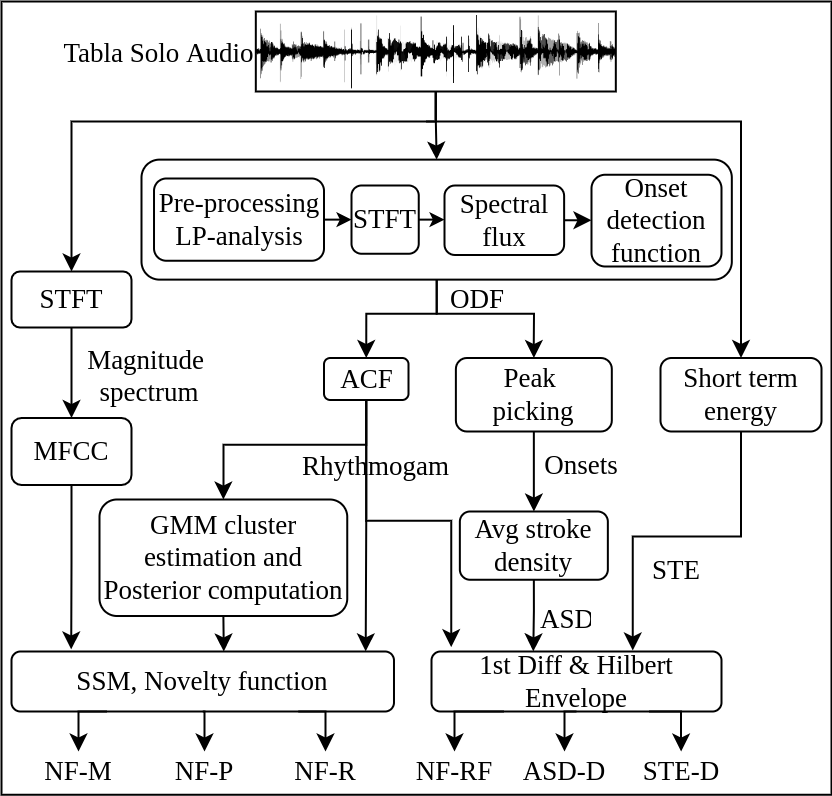}
    }{Figure shows the steps for feature extraction and six different novelty functions.}%
    }
    \caption{Feature Extraction and Novelty Functions Computation Steps. \\ Alt-Text: Figure shows the steps for feature extraction and six different novelty functions.}
    \label{fig:Feature_extraction}
\end{figure}

\begin{figure}[t]
    \centering
    \vspace{-0.2cm}
    \resizebox{\columnwidth}{!}{%
    \pdftooltip[]{
    \includegraphics[trim={3.5cm 0.6cm 3.1cm 1.7cm}, clip]{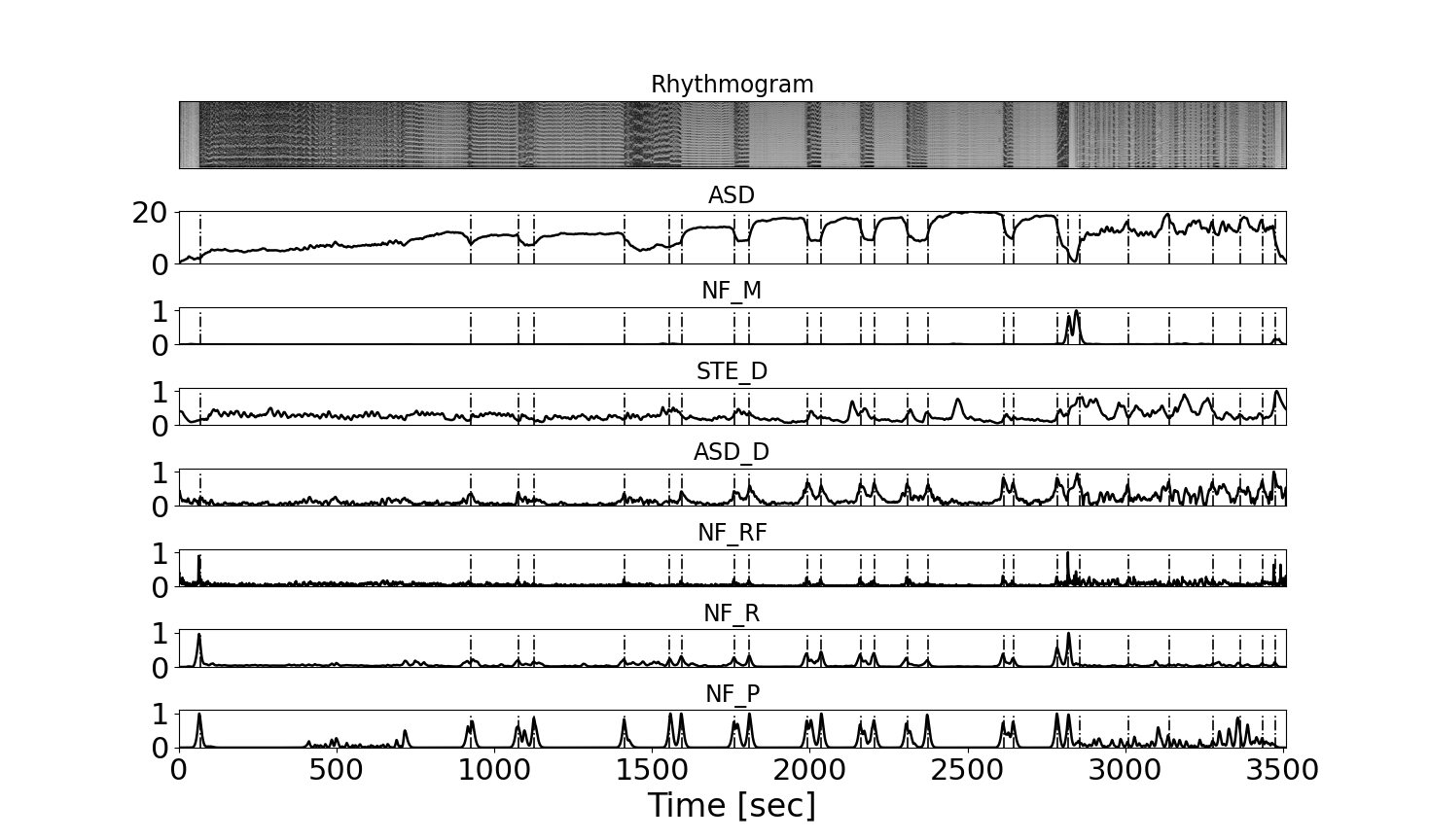}
    }{Figure shows the Rhythmogram, Average Stroke Density (ASD), and different Novelty Functions (NF) Representation for Concert-1. The ground truth segment boundaries are marked by vertical dotted lines.}%
    }
    \caption{Rhythmogram, Average Stroke Density (ASD), and different Novelty Functions (NF) Representation for Concert-1. The ground truth segment boundaries are marked by vertical dotted lines. \\ Alt-Text: Figure shows the Rhythmogram, Average Stroke Density (ASD), and different Novelty Functions (NF) Representation for Concert-1. The ground truth segment boundaries are marked by vertical dotted lines.}
    \label{fig:Rhythmogram_1}
\end{figure}

\begin{figure}[t]
    \centering
    \vspace{-0.2cm}
    \resizebox{\columnwidth}{!}{%
    \pdftooltip[]{
    \includegraphics[trim={3.5cm 0.6cm 3.3cm 1.6cm}, clip]{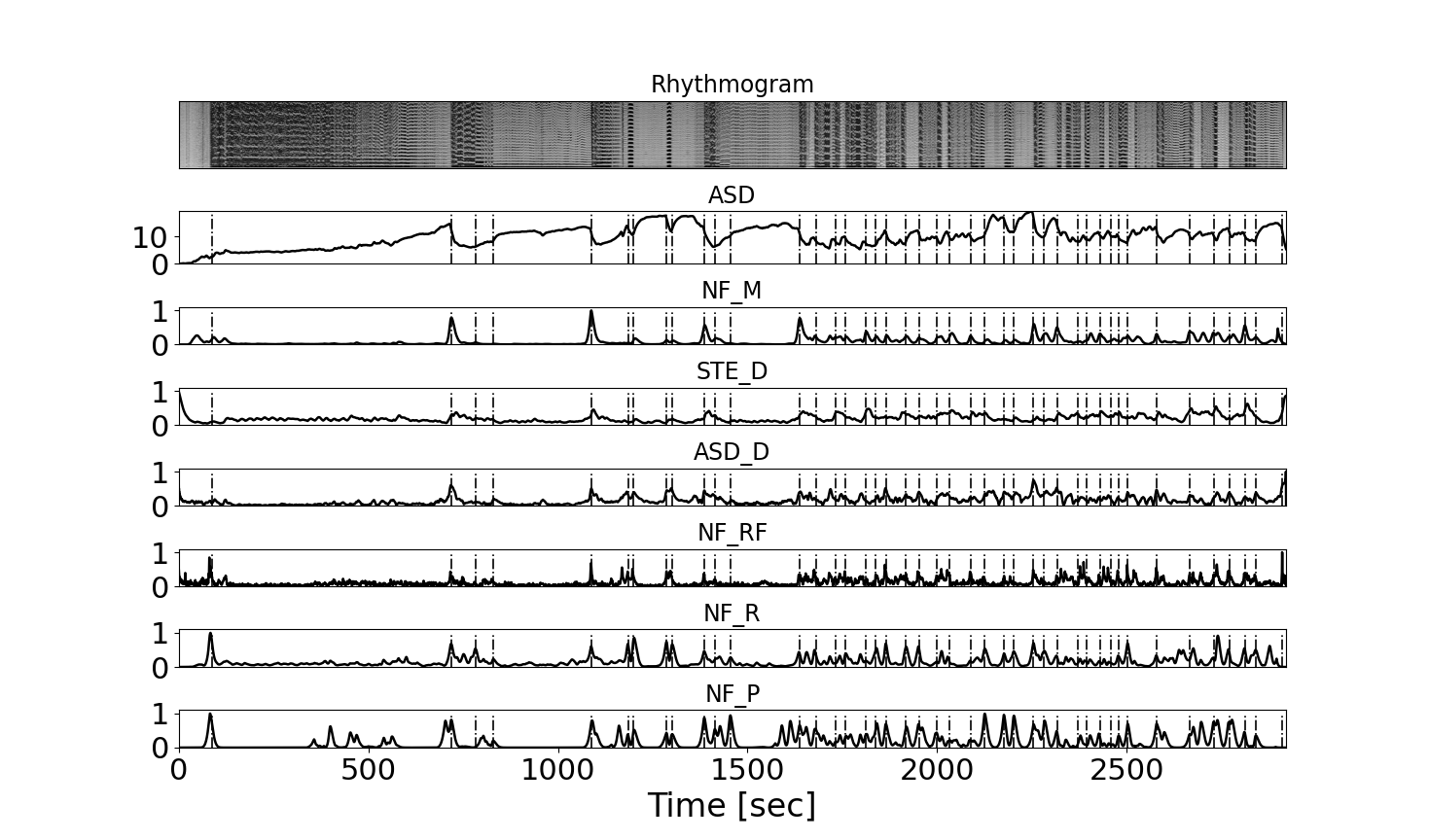}
    }{Figure shows the Rhythmogram, Average Stroke Density (ASD), and different Novelty Functions (NF) Representation for Concert-2. The ground truth segment boundaries are marked by vertical dotted lines.}%
    }
    \caption{Rhythmogram, Average Stroke Density (ASD), and different Novelty Functions (NF) Representation for Concert-2. The ground truth segment boundaries are marked by vertical dotted lines. \\ Alt-Text: Figure shows the Rhythmogram, Average Stroke Density (ASD), and different Novelty Functions (NF) Representation for Concert-2. The ground truth segment boundaries are marked by vertical dotted lines.}
    \label{fig:Rhythmogram_2}
\end{figure}

The steps involved in feature extraction and computing novelty functions are shown in Figure \ref{fig:Feature_extraction}. The 1st difference of ASD (ASD-D), rhythmogram flux (NF-RF), and 1st difference of STE (STE-D) captures the sudden change in tempo, rhythm, and energy, respectively, which can serve as a potential novelty function for the segmentation task. We compute the Hilbert envelope on the difference signal and use it as NF. Hilbert envelope is the magnitude function of complex time function and hence it is unipolar in nature. Figure \ref{fig:Rhythmogram_1} and \ref{fig:Rhythmogram_2} shows different NFs obtained for two different concerts. The ASD-D, NF-RF, and STE-D are observed to be noisy. This is so because the differentiation is a highpass filter, and even the small changes in the features are enhanced in the derivative. The NF-P, NF-R, and NF-M obtained by SSM-kernel convolution are relatively less noisy. This confirms the property of SSM in capturing the homogeneity\cite{verma2015structural}. The homogeneity of the sections is clearly seen in NF-P as well. We can observe that the NF-P is flat around the frames where ASD is not changing much. This verifies that NF-P is indeed capturing the underlying tempo information.


Due to local tempo and timbre variations, peaks are observed around the boundary instances with other spurious peaks. Peak picking is performed on the NF of each feature by maintaining the minimum inter-peak interval of 10 s. Initially, the peaks are picked on individual NFs and considered as boundary instances. We further explore two different ways of getting the boundary instances. The first one is to combine the individual NFs by computing the average and peak picking. The second is to fuse the boundary information from each NFs after the peak picking, which is more flexible in performance tuning.



Each of the novelty functions works well in certain situations. Averaging out different novelty functions worked very well for most concerts. For fusing the detected peaks, NF-P is taken as the reference as it is less noisy. Boundary candidates picked from each rhythm-based NFs (ASD-D, NF-RF, NF-R, NF-P) are fused using a majority decision rule (i.e., the peaks from two or more NFs out of three are examined for coincidence with NF-P as reference).



\subsection{Supervised approaches}

The supervised classifier models are trained on the original and augmented dataset, which includes speed perturbation (see Section \ref{subsec:train_test_eval}). All the frames in $\pm 5$ seconds about the manually labelled boundaries are labelled as boundary frames \cite{ullrich2014boundary}. This accounts for the ambiguities in the manual boundary annotation. The number of non-boundary frames is very high compared to boundary frames. Thus to balance the ratio of boundary and non-boundary frames, only the boundary frames from the augmented audios are retained, along with all of the frames from the original dataset (Data augmentation is explained in Section \ref{subsec:train_test_eval}).


Since the segment boundaries are depicted by a sudden change in the tempo and rhythmic structures, the task benefits from the use of context frames. Thus by keeping the current frame at the center, $\pm C$ adjacent feature frames are included as the context frames forming a window. The target corresponding to each window is a label that indicates if the central frame is a manually labelled boundary frame or not. The probability of the central frame in a window being a boundary frame is estimated at the output of the classifier. Since segment boundary cues can be spread over several frames, predictions within a 5s window are replaced by a single prediction on the frame having the highest probability.



\subsubsection{Random Forest Classifier}

A random forest (RF) is an ensemble of decision trees. The RF classifier outputs the class labels with the majority voting rule on each decision tree prediction. The feature vectors with the current and context frames are fed as the input. Each of the training vectors is assigned the target 1 or 0, indicating if the current frame is a manually labelled boundary or not. The classifier is trained independently on rhythmogram and posterior features. We experimented with and tuned the model hyperparameters, such as the number of trees, and the maximum number of levels in a tree, with and without bootstrap aggregation. The number of decision trees is varied between 10 and 50 in steps of 5, while the context duration (C) is varied from $\pm 5$ to $\pm 20$ seconds in step 5s. The maximum number of levels in the tree is 5 for the rhythmogram and 3 for posteriors inputs.



\subsubsection{CNN Classifier}
\label{subsec:CNN_Classifier}

The rhythmogram is split into smaller overlapping chunks having $\pm C$ context frames and used as the input to the network. All the input data chunks are normalized to have zero mean and unit variance in a single input chunk. The network is trained to learn the frame-level targets. Each target output is 1 or 0, indicating if the center frame has the labelled boundary or not. The boundary predictions are at the frame resolution of 0.5 s.

The CNN network architecture is adapted from \cite{ullrich2014boundary} network with modifications. The model has three convolutional layers and two fully connected layers. Each CNN layer is followed by BatchNormalization \cite{ioffe2015batch} and \emph{ReLU} activation \cite{nair2010rectified}. The horizontal Sobel kernels are used in convolutional layers \cite{Sobel_filter}. The kernel size, number of kernels, and padding of each convolution layer are (100 x 10, 256, 3), (256 x 8, 128, 2), (128 x 4, 64, 1), respectively. This 1D-CNN component configuration is adapted from \cite{oord2018representation}. The model further consists of two fully connected layers with 256 and 2 hidden units, respectively. \emph{Sigmoid} and \emph{softmax} activations are used for fully connected layers. The model is trained using Adam optimizer \cite{kingma2014adam} and binary cross-entropy loss on mini-batches of a size of 32 and a learning rate of 0.01. We have used a learning schedule of dropping the learning rate to half of the current whenever the validation loss doesn't decrease \cite{audhkhasi2017end}. We experimented with different kernels sizes and other activation function combinations — \emph{sigmoid} for the convolutional layers and \emph{ReLU} and with 5\% dropout.

\subsection{Post processing -- merging the segments}

As mentioned in the challenges, small pauses may exist within the rendition of the same compositions. This causes a change in rhythm and tempo structure in those instances. Since our segmentation models inherently depend on sudden tempo and rhythm changes to predict the segment boundaries, these small pauses cause false positives. Thus, we try to look for the tempo and posterior information in the adjacent segments and merge adjacent rhythmically similar segments with the same tempo.

After we get the hypothesized segment boundaries from segmentation systems (Section \ref{Sec4:Segmentation_methods}), we compute the mean value of ASD (M-ASD) in each segment. In Figure \ref{Fig:Pe-Ka-GTC_Boundary}, ASD and M-ASD are plotted using green and black solid curves respectively, for three different concerts. If the M-ASD difference between the adjacent segment is less than three strokes per second, then the posteriors from these segments are considered to decide whether these segments can be merged. The posterior class for each frame of the two segments is checked. The two segments are merged if both the segments have the same posterior classes in the majority of the frames. This also helped in merging smaller \d{t}h\={e}k\={a} segments between the compositions. 

\section{Section Classification and Labeling}
\label{sec5:Section_Classification_and_Labeling}
Once the local segment boundary results are obtained from the segmentation models, the next task is to classify and label the segments appropriately. Segment classification and labeling is the task consisting of: (1) Classifying the concert segments in to high-level sections such as al\={a}p, p\={e}\`{s}k\={a}r (Pe), k\={a}yad\={a} (Ka), and ga\d{t}-tukd\={a}-chakradh\={a}r (GTC). (2) \Gharana\ labeling the individual segments (compositions).



\subsection{Section classification}

Each of these four sections al\={a}p, Pe, Ka, and GTC have unique structural, positional, and duration characteristics which are common across the concerts. We consider these characteristic cues for initial section markings and updates. The characteristic cues for each section are described in Section \ref{subsec:concert_structure}. Since the al\={a}p section does not have percussive strokes and is always present at the start of the concert, the initial segments are checked for the percussive onsets, and the segments which do not contain percussion onsets, are classified as al\={a}p section. The timbre information is very helpful in classifying the percussion and non-percussion (al\={a}p in this case). Thus we also use the MFCC features for verifying the al\={a}p boundary. 

\begin{figure}[t]
    \centering
    \vspace{-0.2cm}
    \resizebox{\columnwidth}{!}{%
    \pdftooltip[]{
    \includegraphics[trim={3.4cm 0.5cm 3.5cm 0.5cm}, clip]{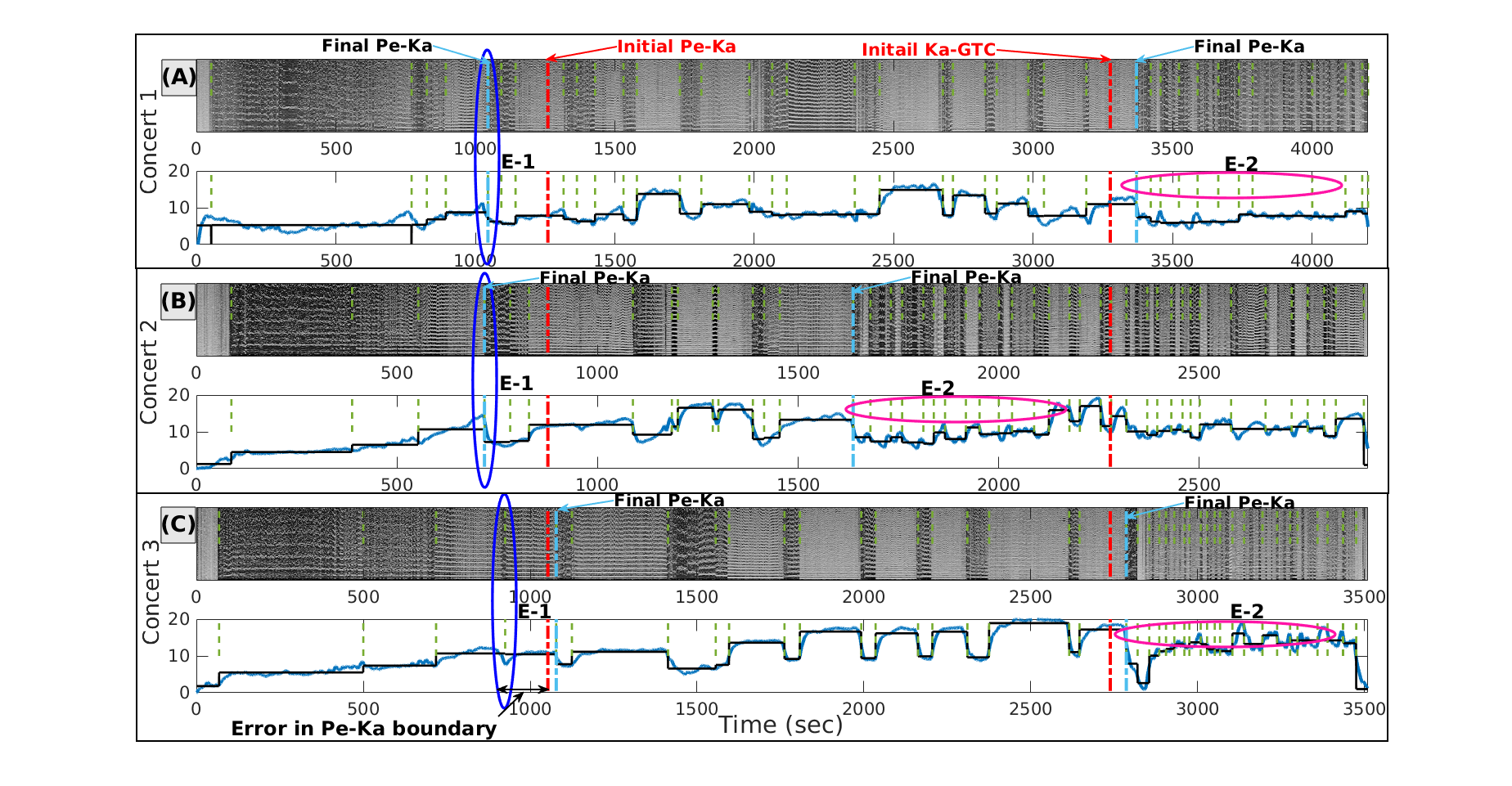}
    }{Figure shows the Section Classification update depiction for three different concerts. \emph{Green dotted lines}: Segmentation boundaries. \emph{Black solid curves}: mean value of ASD (M-ASD) in each segment. \emph{Blue E1}: Ground truth Pe-Ka boundary. \emph{E2}: Start of multiple small segments.}%
    }
    \caption{Section Classification update depiction for three different concerts. \emph{Green dotted lines}: Segmentation boundaries. \emph{Black solid curves}: mean value of ASD (M-ASD) in each segment. \emph{Blue E1}: Ground truth Pe-Ka boundary. \emph{E2}: Start of multiple small segments. \\ Alt-Text: Figure shows the Section Classification update depiction for three different concerts. \emph{Green dotted lines}: Segmentation boundaries. \emph{Black solid curves}: mean value of ASD (M-ASD) in each segment. \emph{Blue E1}: Ground truth Pe-Ka boundary. \emph{E2}: Start of multiple small segments.}
    \label{Fig:Pe-Ka-GTC_Boundary}
\end{figure}

Labeling the next three sections involves initialization and re-estimation of the label boundaries. Figure \ref{Fig:Pe-Ka-GTC_Boundary} illustrates the initial and updated section boundaries for three different tabla solo concerts having different lengths. The p\={e}\`{s}k\={a}r section always occurs at the start of the concert, and the average duration ratio is one-third the length of the concert (Figure \ref{fig:mean_variance_data}). Thus the Pe-Ka boundary is initialized at the instance that is $0.3$ times the length of the concert by taking the ratio of the length of the p\={e}\`{s}k\={a}r section. The k\={a}yad\={a} section is the middle larger section and it occurs after p\={e}\`{s}k\={a}r and before GTC (see Section \ref{sec2:dataset_Description}). The ratio of the average duration of the Ka section to the whole concert is found to be $0.42$. The initial Ka-GTC boundary is fixed at the instance of $0.3 + 0.42 = 0.72$ times the length of the concert. The initialized Pe-Ka, and Ka-GTC boundaries are marked using vertical dotted red lines for all the three concerts in Figure \ref{Fig:Pe-Ka-GTC_Boundary}.

The boundary between al\={a}p and p\={e}\`{s}k\={a}r is accurate in the initial markings itself due to the difference in the timbre and the absence of percussion onsets in al\={a}p. The Pe-Ka and the Ka-GTC boundaries need to be re-aligned considering the cues and musical characteristics. The duration of each segment in GTC sections is very small compared to others (Figure \ref{fig:mean_variance_data} (C)). We check for the repeated smaller segments having a duration of fewer than 50 seconds, and the Ka-GTC boundary is shifted to the start of these smaller segments. These repeated smaller segments are depicted by ellipse E-2, and the updated Pe-Ka boundaries are depicted by vertical blue dotted lines for all the three concerts in Figure \ref{Fig:Pe-Ka-GTC_Boundary}. Each composition is developed and played at a higher speed towards end, followed by a few cycles of \d{t}h\={e}k\={a} at the basic tempo before starting the next composition as described in Section \ref{subsec:concert_structure}. Thus, the ASD increases and decreases, indicating the end of the composition. This sudden drop in ASD followed by \d{t}h\={e}k\={a} in a slower tempo is also seen during the transition from p\={e}\`{s}k\={a}r to k\={a}yad\={a}. We check for this change in ASD around the initially assigned Pe-Ka boundary and update the Pe-Ka boundary as the end of p\={e}\`{s}k\={a}r. The updated Pe-Ka boundaries are depicted by blue vertical dotted lines in Figure \ref{Fig:Pe-Ka-GTC_Boundary}. 




\subsection{\Gharana\ Recognition}
Tabla \gharana\ recognition is a task of identifying the stylistic schools of tabla from tabla recordings. As described in Section \ref{sec2:dataset_Description}, the tabla solo concert consists of several compositions played one after the other. Each composition may be from different \gharana-s. Given the section information such as segment boundaries and section labels (k\={a}yad\={a}, p\={e}\`{s}k\={a}r, GTC), \gharana\ recognition is task is performed only on the k\={a}yad\={a} and GTC sections of the concert, primarily because the ground truth is available only for these two sections. Recognizing the \gharana\ from the p\={e}\`{s}k\={a}r section is not straightforward as it is extempore and the improvisations are usually artiste specific. Table \ref{tab:Gharana_data} shows the details of the annotated data used for \gharana\ recognition task.


\begin{table}
\centering
\caption{Dataset information of \gharana\ labels.}
\begin{tabular}{ccc} 
\hline
\begin{tabular}[c]{@{}c@{}}\textbf{ \Gharana\ Name }\\\textbf{(ID)}\end{tabular} & \begin{tabular}[c]{@{}c@{}}\textbf{No. of}\\\textbf{segments}\end{tabular} & \begin{tabular}[c]{@{}c@{}}\textbf{Durations}\\\textbf{hh:mm:ss}\end{tabular}  \\ 
\hline
Ajrada (\textbf{A})                                                   & 73                                                                         & 2:23:58                                                                        \\
Banaras (\textbf{B})                                                  & 76                                                                         & 3:09:01                                                                        \\
Delhi (\textbf{D})                                                    & 84                                                                         & 2:32:11                                                                        \\
Farrukhabad (\textbf{F})                                              & 96                                                                         & 2:36:50                                                                        \\
Lucknow (\textbf{L})                                                  & 83                                                                         & 2:34:52                                                                        \\
Punjab (\textbf{P})                                                   & 79                                                                         & 2:53:27                                                                        \\
\hline
\end{tabular}
\label{tab:Gharana_data}
\end{table}

Thus, given the segment boundaries, recognizing tabla \gharana\ for each segment (composition) is the third task. Tabla solos are developed, improvised, and elaborated upon a theme through a series of variations according to rhythmic practices \cite{pradhan2011tabla}. Thus some strokes co-occur more often than others in the tabla compositions. Hence it is essential to train the models by encoding the sequence information. The rhythmogram features only have tempo-related information but not sequence-specific language information. Thus we cannot use the ACF features directly for the \gharana\ recognition work. We used the CNN-LSTM model for \gharana\ recognition described in \cite{gowriprasad_r_gharana_ISMIR2021}. The CNN architecture extracts the raw audio's local discriminate features on different stroke sounds. The LSTM networks are trained to classify the \gharana-s by processing the sequence of extracted features from CNNs.

\section{Experiments and Results}
\label{sec6:Experiments_and_Results}


Different experiments are performed addressing different challenges for the three tasks mentioned above. 

\subsection{Train-Test sets and evaluation criteria}
\label{subsec:train_test_eval}

The overall dataset consists of 50 concert recordings with around 2000 ground-truth annotated boundary instances marked as described in Section \ref{sec2:dataset_Description}. We split the 50 concerts into 40-5-5 as train-development-test respectively. The supervised segmentation system evaluation is performed with 10-fold cross-validation. Since different artists use different tonics to perform each concert, we tried to make train-test tonic and artiste mutually exclusive. The train-test tonic is different in each of the seven folds, and the artiste is different in 8 folds. In each fold, the 45 concerts form the train-dev set. Data augmentation is performed to increase the training data diversity. Speed perturbation with the factor of 0.9X and 1.1X (10\% variation) without altering the pitch is done on the entire data using the HPSS-TSM method \cite{6678724, yong2020pytsmod}. The speed perturbation essentially changes the speed and, in turn, changes the rhythmogram structure, thus resulting in a variation in features for training. Each fold has $5 \times 3 = 15$ test concerts, and the final performance is averaged over ten folds (150 concerts). In the case of an unsupervised approach, the validation set is used to tune the hyperparameters, such as the kernel width, and peak picking thresholds. In the case of supervised, it is used to tune the network-related hyperparameters such as number of trees, context window, number of CNN layers, etc.


\subsection{Structural segmentation (Results)}

Structural segmentation in the context of tabla solo involves the detection of start and end instances of composition. Approaching the segmentation task as the boundary detection task, we inspect the presence or absence of a boundary in the uniformly spaced audio frame intervals of 0.5 s. The task is not addressed at the metrical time scale, stroke level but at a larger time scale. Thus the tolerance duration is not in milliseconds as in the case of stroke onset detection \cite{gowriprasad2020onset,bello2005tutorial} but in a larger time scale in "seconds". A detected boundary is considered true-positive if the prediction is within the tolerance of $\pm 5$ seconds of ground truth (GT) boundary; otherwise, it is treated as false-positive. The performance is evaluated using the measures of precision, recall, and weighted-F1 scores.

\subsubsection{Unsupervised Segmentation}

The segmentation evaluation scores like precision, recall, and F-measure for individual novelty functions and their combinations are tabulated in Table \ref{tab:results_unsupervised}. NFs are computed by convolving $(50 \times 50)$ kernel with SSM of different features. Peak picking is performed with an adequate threshold, with a minimum distance between adjacent peaks as 10s. Combo-1 is obtained by averaging the rhythm-based NFs (ASD-D, NF-R, NF-P) by excluding NF-RF, as it found to have larger variance in F-measure. Combo-2 results are obtained by fusing the results of individual NFs after peak picking.


We experimented with smaller kernel sizes such as $(10 \times 10), (20 \times 20)$, resulting in noisy NFs. This decreased the precision as a lot of false positives were reported. Though much larger kernel sizes, such as $(100 \times 100)$, made the NFs smoother, they compromised in resolving the closer boundaries. Thus we used $(50 \times 50)$ kernel owing to the reasons mentioned in Section \ref{subsec:unsupervised_approach}. It is found to produce considerably smooth NFs that result in higher F-measure.

\begin{table}[t]
\centering
\caption{Performance of Unsupervised Segmentation using different feature subsets.}
\begin{tabular}{cccc} 
\toprule
\multirow{2}{*}{\begin{tabular}[c]{@{}c@{}}\textbf{Features}\\\textbf{set}\end{tabular}} & \multicolumn{3}{c}{\textbf{Performance }}                  \\ 
\cline{2-4}
                                                                                         & \textbf{Precision} & \textbf{Recall} & \textbf{F-measure}  \\ 
\hline
NF-M                                                                                     & 0.86               & 0.55            & 0.63                \\
STE-D                                                                                    & 0.77               & 0.81            & 0.78                \\
ASD-D                                                                                    & 0.78               & 0.87            & 0.82                \\
NF-RF                                                                                    & 0.76               & 0.86            & 0.79                \\
NF-R                                                                                     & 0.83               & 0.90            & 0.86                \\
NF-P                                                                                     & 0.80               & \textbf{0.91}   & 0.85                \\
\textbf{Combo-1}                                                                         & 0.86               & 0.90            & \textbf{0.88}       \\
\textbf{Combo-2}                                                                         & \textbf{0.87}      & 0.90            & \textbf{0.88}       \\
\bottomrule
\end{tabular}
\label{tab:results_unsupervised}
\end{table}

Timbre feature MFCC is performing very poorly compared to the rhythm-based features. Since a single tabla is used throughout the concert, the timbre of the tabla does not change throughout. Though the timbre of each stroke is different, the short-term averaging eliminates the stroke combination information within the context window. Even though the stroke combinations vary across the compositions, the stroke set remains the same. We can observe good precision but very poor recall (a lot of false-negative). This indicates that MFCC does not capture the change in rhythm structure. The timbre structure remains more or less the same throughout, making it hard to distinguish between adjacent compositions. Another important factor is that it is common to recite the face theme of the composition often before playing. These are especially seen in the GTC section. This switching between oral recitation and playing causes a drastic switch in timbre features. These drastic changes mask the other timbre changes due to rhythm variations during playing. This resulted in a very low recall.

We can observe that the precision is consistently less than recall in all the other cases, indicating false positives. The change in local rhythm structure, which may be both gradual and abrupt, causes peaks in the novelty function. The gradual change in rhythm structure can be seen often in the p\={e}\`{s}k\={a}r section as it is extempore, and the tempo increases gradually. The abrupt changes due to pauses or inherent compositional characteristics can be seen throughout the concert.

NF-R and NF-P performed better than the rest NFs individually. The NF-R has the highest precision and lowest variance in F-measure while considering individual NF. In the case of NF-R, the entire rhythmogram is used to compute the SSM. The gradual tempo evolution and small pauses are less emphasized than larger changes in rhythmic structure across different compositions. This helped in reducing the false positives. In the case of NF-P, the posteriors are only dimension five, and each dimension captures a different set of tempo-rhythm structures. The gradual change in rhythm structure changes the posteriors at some point. This causes a few false positives in NF-P compared to NF-R. 

Combining the NFs yielded slightly better results compared to the individual NFs. We can observe the increase in precision when the NFs are combined. Each NF highlights the actual boundary instances along with the spurious false positives. One of the advantages was that the spurious peaks are found at different instances in different NFs. Thus, these spurious peaks are de-emphasized, and the actual boundary peak instances common across the NFs are emphasized by taking the ensemble average of NFs (Combo-1). This increased the precision and the overall performance.

A similar behaviour is observed while fusing the information from different novelty functions. The common peaks are picked on the majority voting across the NFs. This helped retain most of the boundary peaks while eliminating the spurious ones, thus increasing the precision. We can observe that the recall is not compromised by these combinations of NFs, which indicates that each considered NF emphasizes the actual boundary instances correctly.

\begin{table}[t]
\centering
\caption{Segmentation Performance of RF Classifier using different feature subset.}
\begin{tabular}{ccccc} 
\toprule
\multirow{2}{*}{\begin{tabular}[c]{@{}c@{}}\textbf{Features}\\\textbf{set}\end{tabular}} & \textbf{Parameters} & \multicolumn{3}{c}{\textbf{Performance }}                  \\ 
\cline{2-5}
                                                                                         & \textbf{\# trees }  & \textbf{Precision} & \textbf{Recall} & \textbf{F-measure}  \\ 
\hline
MFCC                                                                                     & 25                  & 0.37               & 0.59            & 0.45                \\
Rhythmogram                                                                              & 25                  & 0.85               & 0.88            & \textbf{0.87}       \\
Posteriors                                                                               & 30                  & 0.85               & 0.84            & 0.84                \\
\bottomrule
\end{tabular}
\label{tab:results_randomforest}
\end{table}

\subsubsection{Random Forest Classifier}
Table \ref{tab:results_randomforest} presents the segmentation performance for different feature sets using random forest classifier. All the results are reported for the context duration of 10s, which was found to perform best. The performance of MFCC is clearly bad, indicating the inadequacy of timbre features for rhythm-based analysis. It can be observed that the rhythmogram feature alone performs better than the other features. This verifies the consistency of the rhythmogram feature performance as observed in the unsupervised approach.



\subsubsection{CNN - Classifier}

Since rhythmogram features worked best in both the unsupervised and RF methods, we also used the same features for the CNN classifier. We started with the architecture as described in Section \ref{subsec:CNN_Classifier}. We experimented with different kernel sizes for convolution as well as max-pooling layers. We increased the number of convolutional layers, but the performance did not improve, indicating the overfitting of the model. When used in the fully-connected layer, we found that the Relu activation function fetched $1.5\%$ less training loss than the sigmoid. The use of additional max-pooling layers did not affect the performance. The final model architecture fetched the best results and has three convolutional layers and two fully connected layers.

\begin{table}[!h]
\centering
\caption{Segmentation Performance of CNN Classifer on Rhythmogram features.}
\begin{tabular}{ccccc} 
\toprule
\multirow{2}{*}{\textbf{Model}} & \multirow{2}{*}{\begin{tabular}[c]{@{}c@{}}\textbf{Context }\\\textbf{window~\textbf{$C (s)$}}\end{tabular}} & \multicolumn{3}{c}{\textbf{Performance}}                  \\ 
\cline{3-5}
                                &                                                                                                            & \textbf{Precision} & \textbf{Recall} & \textbf{F-measure}  \\ 
\hline
\multirow{3}{*}{CNN-Classifier} & 5                                                                                                          & 0.86               & 0.87            & 0.86                \\
                                & 10                                                                                                         & 0.9                & 0.89            & \textbf{0.89}       \\
                                & 20                                                                                                         & 0.84               & 0.81            & 0.83                \\
\bottomrule
\end{tabular}
\label{tab:CNN_classifier}
\end{table}

Table \ref{tab:CNN_classifier} shows the performance results obtained for different context durations. The context duration values are motivated by the results from the RF classifier. It is evident from the results that increasing the context duration beyond $10s$ affected the performance as a larger context window contains multiple boundaries. This indicates that the larger context window contains multiple changes in rhythmic structure. The context window of 10s performed best in this case.

\subsubsection{Structural Segmentation (Discussion)}

\begin{figure}[t]
    \centering
    \vspace{-0.2cm}
    \resizebox{\columnwidth}{!}{%
    \pdftooltip[]{
    \includegraphics[trim={3cm 1.5cm 4cm 0cm}, clip]{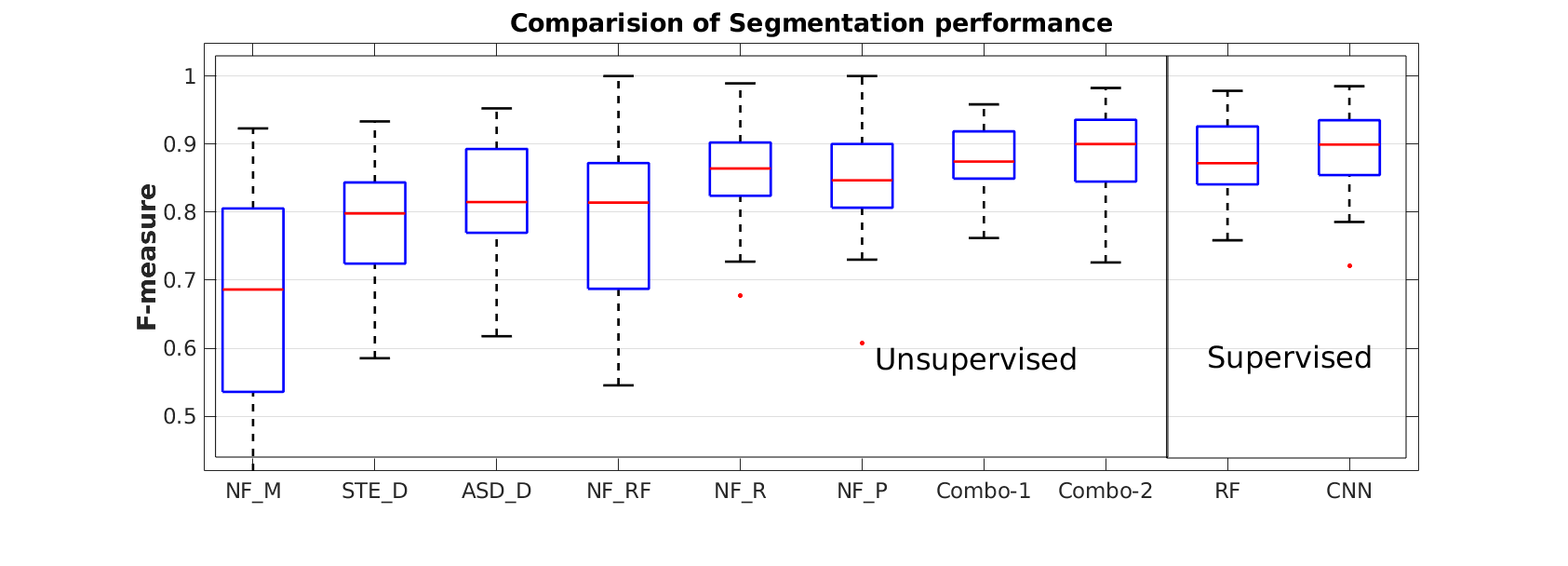}
    }{Figure shows the comparative results for Segmentation task across both supervised and unsupervised methods via boxplots.}%
    }
    \caption{Comparative results for Segmentation task. \\ Alt-Text: Figure shows the comparative results for Segmentation task across both supervised and unsupervised methods via boxplots.}
    \label{fig:Overall_segmentation_comparision}
\end{figure}

Figure \ref{fig:Overall_segmentation_comparision} depicts the performances of different approaches for tabla solo segmentation. Rhythmogram features alone proved to be more efficient in both the supervised and unsupervised approaches. Both the CNN and RF trained on rhythmogram features performed equally well. In the unsupervised approach, combining the peak information and NF-R performed competitively with supervised approach. As mentioned in the challenges section, the short pauses and vocal recitation in the middle of the rendition cause sudden changes in the feature vectors. These cause false positives during segment boundary detection. Thus we can observe that the precision is consistently less than recall almost in all the cases. Merging similar adjacent segments in the post-processing step increased the precision without compromising the recall. Adjacent segments are merged only if they have the same posteriors. Since a single composition is played at multiple speeds, the adjacent segments within the composition do not have the same posteriors, failing to merge and leading to false positives. The precision is relatively less than recall.

Table \ref{tab:individual_section_results} presents the best performing unsupervised and supervised results for the individual sections separately. We can observe that all the models' performance is high on k\={a}yad\={a} sections and low in the GTC section. This is evident as the k\={a}yad\={a}-s are the elaborate compositions explored on a theme and played longer than the GTC compositions. In the case of GTC, the compositions are smaller in length, and the rhythm structure changes rapidly over time, causing false positives. There are many instances where the multiple ga\d{t}-s and tukd\={a}-s are played continuously without any gaps. This causes false negatives. In the case of p\={e}\`{s}k\={a}r section, the tempo, and the rhythmic evolution are gradual. This change also evokes false positives in certain instances, even in the case of posterior features. This can be visualized in Figure \ref{fig:Rhythmogram_1} and Figure \ref{fig:Rhythmogram_2}.



\begin{table}
\centering
\caption{Segmentation performance evaluated on individual section.}
\begin{tabular}{c|ccc|ccc} 
\toprule
\multirow{2}{*}{\textbf{Section}} & \multicolumn{3}{c|}{\textbf{Unsup Combo-2}}               & \multicolumn{3}{c}{\textbf{CNN}}                           \\ 
\cline{2-7}
                                  & \textbf{Precision} & \textbf{Recall} & \textbf{F-measure} & \textbf{Precision} & \textbf{Recall} & \textbf{F-measure}  \\ 
\hline
p\={e}\`{s}k\={a}r                         & 0.69               & \textbf{1}      & 0.81               & 0.71               & \textbf{1}      & 0.83                \\
k\={a}yad\={a}                          & 0.94               & 0.98            & \textbf{0.96}      & 0.94               & 0.95            & \textbf{0.94}       \\
GTC                               & 0.84               & 0.81            & 0.82               & 0.85               & 0.80            & 0.82                \\
\bottomrule
\end{tabular}
\label{tab:individual_section_results}
\end{table}


\subsubsection{Out-of-Domain test data evaluation}
The best performing supervised, and unsupervised systems are further evaluated on five new live concert recordings. These five concerts are considered the out-of-domain test data and, therefore, not used in systems' hyperparameter tuning. Each of these five concerts is about 40-45 minutes in duration and contains more than a hundred annotated boundaries in total. These five concerts are from three different artistes who are distinct from the fifty concert artiste from the in-domain data. Thus the approach and the sequence of compositions played are reasonably different in these five concerts. These five concerts have only segment boundary and Pe-Ka-GTC annotations and do not have \gharana\ labels.

Table \ref{tab:unseen_results} presents the performance of supervised and unsupervised models on the five out-of-domain test concerts. The Combo-2 in the unsupervised approach performed slightly better than the supervised approach. Despite concert and artiste-specific variations, the CNN and RF performance did not degrade much. This indicates the models are not overfitting. The performance on the out-of-domain data validates the efficiency of rhythmogram features. The advantage of rhythmogram features is that it is independent of the different tonic values in the concert, as it is computed only on the onset instances of the strokes. Thus the variations of the concert-specific parameters such as \emph{lehra} (sarangi/harmonium) and tabla tonic variations do not affect the structure of the rhythmogram. We can also observe that only precision degraded while the recall did not. These five recordings are from live concert renditions and have many pauses and vocal recitations in the middle. These characteristic fluctuations resulted in false positives, which in turn decreased the precision.




\begin{table}
\centering
\caption{Segmentation performance on out-of-domain test data (five unseen test concerts).}
\begin{tabular}{ccccc} 
\toprule
\multirow{2}{*}{}      & \multirow{2}{*}{\textbf{Methods}} & \multicolumn{3}{c}{\textbf{Performance }}                  \\ 
\cline{3-5}
                       &                                   & \textbf{Precision} & \textbf{Recall} & \textbf{F-measure}  \\ 
\hline
\multirow{4}{*}{UnSup} & NF-R                              & 0.81               & 0.86            & 0.84                \\
                       & NF-P                              & 0.80               & 0.86            & 0.83                \\
                       & Combo-1                           & 0.85               & 0.86            & 0.85                \\
                       & Combo-2                           & 0.85               & 0.86            & \textbf{0.86}       \\ 
\hline
\multirow{2}{*}{Sup}   & Random Forest                     & 0.79               & 0.87            & 0.83                \\
                       & CNN                               & 0.80               & 0.88            & 0.845               \\
\bottomrule
\end{tabular}
\label{tab:unseen_results}
\end{table}

\subsection{Section classification}

Section classification does not require training, as the algorithm for the task is a set of rules that makes the decision. Thus the classification performance evaluation is done on the entire data. Considering the ground truth annotations, we quantify the performance of section classification by calculating the number of miss-classified frames in the whole recording. We present the evaluation in terms of accuracy: the ratio of the number of frames correctly classified to the total number of frames in the concert. The weighted average of accuracy considering the length of the concert is 92\%. That is, given one hour of segmented concert details, around 55 minutes of the concert frames are correctly labeled as p\={e}\`{s}k\={a}r -- k\={a}yad\={a} -- GTC. The ground truth location for Pe-Ka boundary is marked by ellipse E-1 in Figure \ref{Fig:Pe-Ka-GTC_Boundary}. In the case of concerts (A) and (B), the final Pe-Ka boundary is found to coincide with the E-1. In the case of concert (C), an error in the final Pe-Ka boundary is observed, which is not coinciding with E-1. The frames belonging to Ka, which are misclassified, are also marked in Figure \ref{Fig:Pe-Ka-GTC_Boundary}-C. The section classification accuracy in the case of out-of-domain test concerts is found to be 86\%. This is because the sequence of compositions in the five concerts is slightly different. There were a few instances of ga\d{t}-s and tukd\={a}-s in between the k\={a}yad\={a}-s which is significantly different from that present in the training data.


\subsection{\Gharana\ recognition}

The task of \gharana\ recognition is performed using the CNN-LSTM model described in \cite{gowriprasad_r_gharana_ISMIR2021}. The train-test data split is the same as the segmentation task. The audios from the Ka and GTC sections of the 45 concert audios are used for training, and five concert audios for testing. The section audios are segmented into fifteen seconds smaller chunks with a five seconds overlap. Each fifteen-second chunk is treated as an individual data point for training the network.

\begin{table}
\centering
\caption{\Gharana\ recognition performance results.}
\begin{tabular}{ccc} 
\toprule
\multicolumn{3}{c}{\textbf{\textbf{Performance}}}                        \\ 
\hline
\textbf{\textbf{Experiment}} & \textbf{Weighted F1} & \textbf{Accuracy}  \\ 
\hline
Chunk wise                   & 0.72                 & 0.71               \\
Section wise                 & 0.78                 & 0.79               \\
\bottomrule
\end{tabular}
\label{tab:Gharana_results}
\end{table}

\begin{figure}[t]
    \centering
    \resizebox{\columnwidth}{!}{%
    \pdftooltip[]{
    \includegraphics[trim={3.2cm 0cm 1.6cm 0cm}, clip]{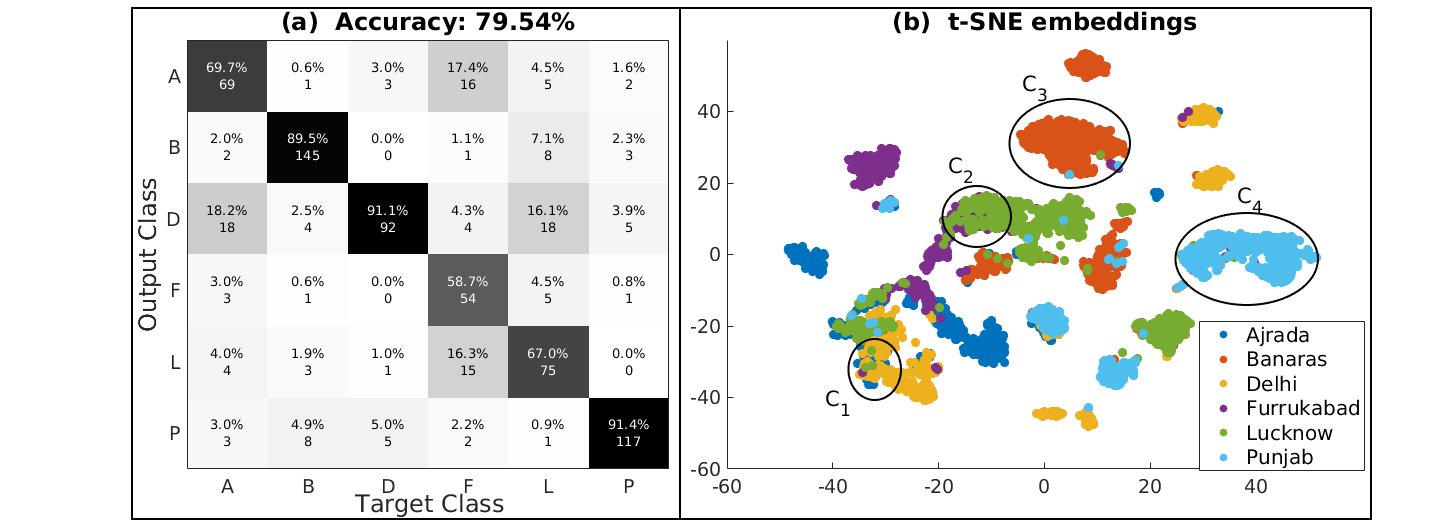}
    }{Figure shows (a) \Gharana\ Recognition Confusion Matrix, (b) t-SNE visualization of test data embedding extracted from CNN-LSTM Model.}%
    }
    \caption{(a) \Gharana\ Recognition Confusion Matrix, (b) t-SNE visualization of test data embedding extracted from CNN-LSTM Model.\\ Alt-Text: Figure shows (a) \Gharana\ Recognition Confusion Matrix, (b) t-SNE visualization of test data embedding extracted from CNN-LSTM Model.}
    \label{Fig:Confusio_matrix_gharana}
\end{figure}

During testing, \gharana\ prediction is made independently on each of the fifteen-second chunks in a particular segment. The number of chunks predicted for each \gharana\ in the segment is counted. The final \gharana\ label for a segment is assigned according to the majority vote. The \gharana\ for the whole segment is labelled by the voting rule, i.e., the \gharana\ to which a maximum number of chunks are predicted. The experiment is 5-fold cross-validated, and the average performance results are tabulated in Table \ref{tab:Gharana_results}. One can observe that the performance is better when evaluated segment-wise. Though a few chunks in a segment confused and miss-classified, most of the chunks in the segment were classified correctly, making the \gharana\ recognition better segment-wise. The \gharana\ recognition task is performed by labeling the automatically segmented audios in this work. The train-split task was kept the same as before (segmentation task). Since the train and test concerts are different in each fold of evaluation, the tonic and artiste of the train-test were maintained distinct. Thus the \gharana\ recognition results are in the context of inter-artiste and inter-tonic conditions as mentioned in \cite{gowriprasad_r_gharana_ISMIR2021}.

The confusion matrix depicting the \gharana\ recognition performance and the t-SNE visualization of the test data embeddings from the penultimate layer of the CNN-LSTM network for fifteen seconds chunk duration are shown in Figure \ref{Fig:Confusio_matrix_gharana}. Different clusters are marked for the benefit of reference. We can observe confusion between Ajrada-Delhi ($C_1$), as well Lucknow-Farrukhabad ($C_2$). This confusion is defensible as the Farrukhabad and Ajrada \gharana-s are developed from Lucknow and Delhi \gharana-s respectively \cite{gottlieb1993solo}. Less confusion and high recognition accuracy are observed in the case of Banaras ($C_3$) and Punjab \gharana\ ($C_4$) embeddings. This is apparent as the Punjab \gharana\ has had an individual existence and is distinctive compared to the other ones \cite{pradhan2011tabla}. Pakh\={a}waj syllables highly influence the Banaras \gharana\ \cite{gottlieb1993solo}, and the technique of playing the bass drum is also unique in Banaras \gharana\ \cite{pradhan2011tabla}. Thus one can verify some of the traditional similarities and differences across the \gharana-s as portrayed and described in the sources \cite{pradhan2011tabla,saxena2006art,gottlieb1993solo} from the obtained results.

\section{Conclusions}
\label{sec7:Conclusion}

This work has addressed an unexplored problem, structural segmentation and labeling of tabla solo concerts. We described the different sections of tabla solo concerts and analyzed their diversity using various statistical parameters. The major contributions of this work are as follows: (i) curating a diverse dataset of tabla solo recordings having section boundary information along with \gharana\ annotated section labels, (ii) evaluating the existing MIR techniques for a musicologically important task segmentation and labeling of tabla solo concert, (iii) formulating Average Stroke Density ASD feature (a representative of surface tempo), which is robust to tempo octave errors, (iv) formulating the class-conditional probability features from the rhythmogram, and (v) exploring the combination of different NFs obtained from different features.

We described the challenges that are unique to the above tasks. We evaluated the performance of unsupervised and supervised methods for the structural segmentation task. The CNN classifier trained on hand-crafted rhythm features performed well on the in-domain dataset with an F-measure of $0.89$ and $0.845$ on the out-of-domain test data. The unsupervised approach based on rhythm features performed competitively on in-domain and out-of-domain datasets, achieving F-measures of $0.88$ and $0.86$, respectively. We found that the rhythmogram is robust to concert and artiste variations. By choosing a suitable combination of hand-crafted features, it appears that the unsupervised approach performs on par with the supervised machine learning approaches.

The labeling task consists of classifying the segments into high-level sections and recognizing the \gharana-s of each segment. As shown in the Section \ref{sec6:Experiments_and_Results}, the results of segment classification and \gharana\ recognition tasks are reliable with the performance accuracies of $92\%$ and $79\%$, respectively. The outcome of the tasks, such as section boundaries, section-specific, and \gharana\ specific labeling, are vital metadata for analysis of tabla solo performances.

\section{Data Availability}
Manually annotated metadata of all the audio files and the information about the audios sources in the dataset will be provided upon request.

\section{Acknowledgments}
\label{sec:Acknowledgments}

This work was greatly backed up by the constant guidance of Tabla master Sri. Ramesh Dhannur, Sri. Aneesh Pradhan and Sri. Kiran Yavagal. The authors are grateful to Dr. Ajay Srinivasamurthy for his support, and timely advice for the work.

\section*{Glossary}

\begin{description}

\item [t\={a}l]The rhythmic framework of Hindustani music.
\item [al\={a}p] An unmetered melodic improvisation.
\item [\d{t}h\={e}k\={a}]The basic stroke pattern associated with a t\={a}l.
\item [ghar\={a}n\={a}]The stylistic schools of Hindustani music.
\item [m\={a}tra]The lowest defined metrical pulse in Hindustani music (equivalent to a beat).
\item [sam] The first m\={a}tra of rhythm cycle.
\item [pakh\={a}waj] A double barrel drum used as rhythm accompaniment in Hindustani music.
\item [lehra] Melodic accompaniment in tabla solo performances which keeps track of the metric tempo.
\end{description}

\medskip

\bibliography{reference.bib} 

\begin{thebibliography}{}

\bibitem [\protect \citeauthoryear {%
A.%
, Bhattacharjee%
\BCBL {}\ \BBA {} Rao%
}{%
A.%
\ \protect \BOthers {.}}{%
{\protect \APACyear {2021}}%
}]{%
rohit_ismir_2021}
\APACinsertmetastar {%
rohit_ismir_2021}%
\begin{APACrefauthors}%
A., R\BPBI M.%
, Bhattacharjee, A.%
\BCBL {}\ \BBA {} Rao, P.%
\end{APACrefauthors}%
\unskip\
\newblock
\APACrefYearMonthDay{2021}{}{}.
\newblock
{\BBOQ}\APACrefatitle {Four-way Classification of Tabla Strokes with Models
  Adapted from Automatic Drum Transcription} {Four-way classification of tabla
  strokes with models adapted from automatic drum transcription}.{\BBCQ}
\newblock
\BIn{} \APACrefbtitle {Proceedings of the 22nd International Society for Music
  Information Retrieval Conference, ISMIR 2021, Online, November 7-12, 2021}
  {Proceedings of the 22nd international society for music information
  retrieval conference, ismir 2021, online, november 7-12, 2021}\ (\BPG~19-26).
\newblock
\begin{APACrefURL} \url{https://archives.ismir.net/ismir2021/paper/000001.pdf}
  \end{APACrefURL}
\PrintBackRefs{\CurrentBib}

\bibitem [\protect \citeauthoryear {%
Allegraud%
\ \protect \BOthers {.}}{%
Allegraud%
\ \protect \BOthers {.}}{%
{\protect \APACyear {2019}}%
}]{%
allegraud2019learning}
\APACinsertmetastar {%
allegraud2019learning}%
\begin{APACrefauthors}%
Allegraud, P.%
, Bigo, L.%
, Feisthauer, L.%
, Giraud, M.%
, Groult, R.%
, Leguy, E.%
\BCBL {}\ \BBA {} Lev{\'e}, F.%
\end{APACrefauthors}%
\unskip\
\newblock
\APACrefYearMonthDay{2019}{}{}.
\newblock
{\BBOQ}\APACrefatitle {Learning Sonata Form Structure on Mozart's String
  Quartets} {Learning sonata form structure on mozart's string
  quartets}.{\BBCQ}
\newblock
\APACjournalVolNumPages{Transactions of the International Society for Music
  Information Retrieval (TISMIR)}{2}{1}{82--96}.
\PrintBackRefs{\CurrentBib}

\bibitem [\protect \citeauthoryear {%
Anantapadmanabhan%
, Bello%
, Krishnan%
\BCBL {}\ \BBA {} Murthy%
}{%
Anantapadmanabhan%
\ \protect \BOthers {.}}{%
{\protect \APACyear {2014}}%
}]{%
anantapadmanabhan2014tonic}
\APACinsertmetastar {%
anantapadmanabhan2014tonic}%
\begin{APACrefauthors}%
Anantapadmanabhan, A.%
, Bello, J.%
, Krishnan, R.%
\BCBL {}\ \BBA {} Murthy, H.%
\end{APACrefauthors}%
\unskip\
\newblock
\APACrefYearMonthDay{2014}{}{}.
\newblock
{\BBOQ}\APACrefatitle {Tonic-independent stroke transcription of the mridangam}
  {Tonic-independent stroke transcription of the mridangam}.{\BBCQ}
\newblock
\BIn{} \APACrefbtitle {Audio Engineering Society Conference: 53rd International
  Conference: Semantic Audio.} {Audio engineering society conference: 53rd
  international conference: Semantic audio.}
\PrintBackRefs{\CurrentBib}

\bibitem [\protect \citeauthoryear {%
Anantapadmanabhan%
, Bellur%
\BCBL {}\ \BBA {} Murthy%
}{%
Anantapadmanabhan%
\ \protect \BOthers {.}}{%
{\protect \APACyear {2013}}%
}]{%
anantapadmanabhan2013modal}
\APACinsertmetastar {%
anantapadmanabhan2013modal}%
\begin{APACrefauthors}%
Anantapadmanabhan, A.%
, Bellur, A.%
\BCBL {}\ \BBA {} Murthy, H\BPBI A.%
\end{APACrefauthors}%
\unskip\
\newblock
\APACrefYearMonthDay{2013}{}{}.
\newblock
{\BBOQ}\APACrefatitle {Modal analysis and transcription of strokes of the
  mridangam using non-negative matrix factorization} {Modal analysis and
  transcription of strokes of the mridangam using non-negative matrix
  factorization}.{\BBCQ}
\newblock
\BIn{} \APACrefbtitle {IEEE International Conference on Acoustics, Speech and
  Signal Processing, 2013.} {Ieee international conference on acoustics, speech
  and signal processing, 2013.}
\PrintBackRefs{\CurrentBib}

\bibitem [\protect \citeauthoryear {%
Audhkhasi%
, Rosenberg%
, Sethy%
, Ramabhadran%
\BCBL {}\ \BBA {} Kingsbury%
}{%
Audhkhasi%
\ \protect \BOthers {.}}{%
{\protect \APACyear {2017}}%
}]{%
audhkhasi2017end}
\APACinsertmetastar {%
audhkhasi2017end}%
\begin{APACrefauthors}%
Audhkhasi, K.%
, Rosenberg, A.%
, Sethy, A.%
, Ramabhadran, B.%
\BCBL {}\ \BBA {} Kingsbury, B.%
\end{APACrefauthors}%
\unskip\
\newblock
\APACrefYearMonthDay{2017}{}{}.
\newblock
{\BBOQ}\APACrefatitle {End-to-end ASR-free keyword search from speech}
  {End-to-end asr-free keyword search from speech}.{\BBCQ}
\newblock
\APACjournalVolNumPages{IEEE Journal of Selected Topics in Signal
  Processing}{11}{8}{1351--1359}.
\PrintBackRefs{\CurrentBib}

\bibitem [\protect \citeauthoryear {%
Bagchee%
}{%
Bagchee%
}{%
{\protect \APACyear {1998}}%
}]{%
bagchee1998nad}
\APACinsertmetastar {%
bagchee1998nad}%
\begin{APACrefauthors}%
Bagchee, S.%
\end{APACrefauthors}%
\unskip\
\newblock
\APACrefYear{1998}.
\newblock
\APACrefbtitle {N{\=a}d: Understanding r{\=a}ga music} {N{\=a}d: Understanding
  r{\=a}ga music}.
\newblock
\APACaddressPublisher{}{Eeshwar}.
\PrintBackRefs{\CurrentBib}

\bibitem [\protect \citeauthoryear {%
Bartsch%
\ \BBA {} Wakefield%
}{%
Bartsch%
\ \BBA {} Wakefield%
}{%
{\protect \APACyear {2005}}%
}]{%
bartsch2005audio}
\APACinsertmetastar {%
bartsch2005audio}%
\begin{APACrefauthors}%
Bartsch, M\BPBI A.%
\BCBT {}\ \BBA {} Wakefield, G\BPBI H.%
\end{APACrefauthors}%
\unskip\
\newblock
\APACrefYearMonthDay{2005}{}{}.
\newblock
{\BBOQ}\APACrefatitle {Audio thumbnailing of popular music using chroma-based
  representations} {Audio thumbnailing of popular music using chroma-based
  representations}.{\BBCQ}
\newblock
\APACjournalVolNumPages{IEEE Transactions on multimedia}{7}{1}{96--104}.
\PrintBackRefs{\CurrentBib}

\bibitem [\protect \citeauthoryear {%
Bello%
\ \protect \BOthers {.}}{%
Bello%
\ \protect \BOthers {.}}{%
{\protect \APACyear {2005}}%
}]{%
bello2005tutorial}
\APACinsertmetastar {%
bello2005tutorial}%
\begin{APACrefauthors}%
Bello, J\BPBI P.%
, Daudet, L.%
, Abdallah, S.%
, Duxbury, C.%
, Davies, M.%
\BCBL {}\ \BBA {} Sandler, M\BPBI B.%
\end{APACrefauthors}%
\unskip\
\newblock
\APACrefYearMonthDay{2005}{}{}.
\newblock
{\BBOQ}\APACrefatitle {A tutorial on onset detection in music signals} {A
  tutorial on onset detection in music signals}.{\BBCQ}
\newblock
\APACjournalVolNumPages{IEEE Transactions on speech and audio
  processing}{13}{5}{1035--1047}.
\PrintBackRefs{\CurrentBib}

\bibitem [\protect \citeauthoryear {%
Chordia%
}{%
Chordia%
}{%
{\protect \APACyear {2005}}%
}]{%
chordia2005segmentation}
\APACinsertmetastar {%
chordia2005segmentation}%
\begin{APACrefauthors}%
Chordia, P.%
\end{APACrefauthors}%
\unskip\
\newblock
\APACrefYearMonthDay{2005}{}{}.
\newblock
{\BBOQ}\APACrefatitle {Segmentation and Recognition of Tabla Strokes.}
  {Segmentation and recognition of tabla strokes.}{\BBCQ}
\newblock
\BIn{} \APACrefbtitle {Proc. 6th International Society for Music Information
  Retrieval (ISMIR), 2005.} {Proc. 6th international society for music
  information retrieval (ismir), 2005.}
\PrintBackRefs{\CurrentBib}

\bibitem [\protect \citeauthoryear {%
Chordia%
, Sastry%
, Mallikarjuna%
\BCBL {}\ \BBA {} Albin%
}{%
Chordia%
\ \protect \BOthers {.}}{%
{\protect \APACyear {2010}}%
}]{%
chordia2010multiple}
\APACinsertmetastar {%
chordia2010multiple}%
\begin{APACrefauthors}%
Chordia, P.%
, Sastry, A.%
, Mallikarjuna, T.%
\BCBL {}\ \BBA {} Albin, A.%
\end{APACrefauthors}%
\unskip\
\newblock
\APACrefYearMonthDay{2010}{}{}.
\newblock
{\BBOQ}\APACrefatitle {Multiple Viewpoints Modeling of Tabla Sequences.}
  {Multiple viewpoints modeling of tabla sequences.}{\BBCQ}
\newblock
\BIn{} \APACrefbtitle {Proc. 11th International Society for Music Information
  Retrieval (ISMIR), 2010} {Proc. 11th international society for music
  information retrieval (ismir), 2010}\ (\BPG~11th).
\PrintBackRefs{\CurrentBib}

\bibitem [\protect \citeauthoryear {%
Chordia%
, Sastry%
\BCBL {}\ \BBA {} {\c{S}}ent{\"u}rk%
}{%
Chordia%
\ \protect \BOthers {.}}{%
{\protect \APACyear {2011}}%
}]{%
chordia2011predictive}
\APACinsertmetastar {%
chordia2011predictive}%
\begin{APACrefauthors}%
Chordia, P.%
, Sastry, A.%
\BCBL {}\ \BBA {} {\c{S}}ent{\"u}rk, S.%
\end{APACrefauthors}%
\unskip\
\newblock
\APACrefYearMonthDay{2011}{}{}.
\newblock
{\BBOQ}\APACrefatitle {Predictive tabla modelling using variable-length Markov
  and hidden Markov models} {Predictive tabla modelling using variable-length
  markov and hidden markov models}.{\BBCQ}
\newblock
\APACjournalVolNumPages{Journal of New Music Research}{40}{2}{105--118}.
\PrintBackRefs{\CurrentBib}

\bibitem [\protect \citeauthoryear {%
Cooper%
\ \BBA {} Foote%
}{%
Cooper%
\ \BBA {} Foote%
}{%
{\protect \APACyear {2003}}%
}]{%
cooper2003summarizing}
\APACinsertmetastar {%
cooper2003summarizing}%
\begin{APACrefauthors}%
Cooper, M.%
\BCBT {}\ \BBA {} Foote, J.%
\end{APACrefauthors}%
\unskip\
\newblock
\APACrefYearMonthDay{2003}{}{}.
\newblock
{\BBOQ}\APACrefatitle {Summarizing popular music via structural similarity
  analysis} {Summarizing popular music via structural similarity
  analysis}.{\BBCQ}
\newblock
\BIn{} \APACrefbtitle {2003 IEEE Workshop on Applications of Signal Processing
  to Audio and Acoustics (IEEE Cat. No. 03TH8684)} {2003 ieee workshop on
  applications of signal processing to audio and acoustics (ieee cat. no.
  03th8684)}\ (\BPGS\ 127--130).
\PrintBackRefs{\CurrentBib}

\bibitem [\protect \citeauthoryear {%
Dannenberg%
\ \BBA {} Goto%
}{%
Dannenberg%
\ \BBA {} Goto%
}{%
{\protect \APACyear {2008}}%
}]{%
dannenberg2008music}
\APACinsertmetastar {%
dannenberg2008music}%
\begin{APACrefauthors}%
Dannenberg, R\BPBI B.%
\BCBT {}\ \BBA {} Goto, M.%
\end{APACrefauthors}%
\unskip\
\newblock
\APACrefYearMonthDay{2008}{}{}.
\newblock
{\BBOQ}\APACrefatitle {Music structure analysis from acoustic signals} {Music
  structure analysis from acoustic signals}.{\BBCQ}
\newblock
\BIn{} \APACrefbtitle {Handbook of signal processing in acoustics} {Handbook of
  signal processing in acoustics}\ (\BPGS\ 305--331).
\newblock
\APACaddressPublisher{}{Springer}.
\PrintBackRefs{\CurrentBib}

\bibitem [\protect \citeauthoryear {%
Dixon%
}{%
Dixon%
}{%
{\protect \APACyear {2006}}%
}]{%
dixon2006simple}
\APACinsertmetastar {%
dixon2006simple}%
\begin{APACrefauthors}%
Dixon, S.%
\end{APACrefauthors}%
\unskip\
\newblock
\APACrefYearMonthDay{2006}{}{}.
\newblock
{\BBOQ}\APACrefatitle {Simple spectrum-based onset detection} {Simple
  spectrum-based onset detection}.{\BBCQ}
\newblock
\APACjournalVolNumPages{MIREX 2006}{}{}{62}.
\PrintBackRefs{\CurrentBib}

\bibitem [\protect \citeauthoryear {%
Driedger%
, Müller%
\BCBL {}\ \BBA {} Ewert%
}{%
Driedger%
\ \protect \BOthers {.}}{%
{\protect \APACyear {2014}}%
}]{%
6678724}
\APACinsertmetastar {%
6678724}%
\begin{APACrefauthors}%
Driedger, J.%
, Müller, M.%
\BCBL {}\ \BBA {} Ewert, S.%
\end{APACrefauthors}%
\unskip\
\newblock
\APACrefYearMonthDay{2014}{}{}.
\newblock
{\BBOQ}\APACrefatitle {Improving Time-Scale Modification of Music Signals Using
  Harmonic-Percussive Separation} {Improving time-scale modification of music
  signals using harmonic-percussive separation}.{\BBCQ}
\newblock
\APACjournalVolNumPages{IEEE Signal Processing Letters}{21}{1}{105-109}.
\newblock
\begin{APACrefDOI} \doi{10.1109/LSP.2013.2294023} \end{APACrefDOI}
\PrintBackRefs{\CurrentBib}

\bibitem [\protect \citeauthoryear {%
Foote%
}{%
Foote%
}{%
{\protect \APACyear {2000}}%
}]{%
foote2000automatic}
\APACinsertmetastar {%
foote2000automatic}%
\begin{APACrefauthors}%
Foote, J.%
\end{APACrefauthors}%
\unskip\
\newblock
\APACrefYearMonthDay{2000}{}{}.
\newblock
{\BBOQ}\APACrefatitle {Automatic audio segmentation using a measure of audio
  novelty} {Automatic audio segmentation using a measure of audio
  novelty}.{\BBCQ}
\newblock
\BIn{} \APACrefbtitle {2000 ieee international conference on multimedia and
  expo. icme2000. proceedings. latest advances in the fast changing world of
  multimedia (cat. no. 00th8532)} {2000 ieee international conference on
  multimedia and expo. icme2000. proceedings. latest advances in the fast
  changing world of multimedia (cat. no. 00th8532)}\ (\BVOL~1, \BPGS\
  452--455).
\PrintBackRefs{\CurrentBib}

\bibitem [\protect \citeauthoryear {%
Gillet%
\ \BBA {} Richard%
}{%
Gillet%
\ \BBA {} Richard%
}{%
{\protect \APACyear {October 2003}}%
}]{%
gillet2003automatic}
\APACinsertmetastar {%
gillet2003automatic}%
\begin{APACrefauthors}%
Gillet, O.%
\BCBT {}\ \BBA {} Richard.%
\end{APACrefauthors}%
\unskip\
\newblock
\APACrefYearMonthDay{October 2003}{}{}.
\newblock
{\BBOQ}\APACrefatitle {Automatic labelling of tabla signals} {Automatic
  labelling of tabla signals}.{\BBCQ}
\newblock
\BIn{} \APACrefbtitle {Proc. 4th International Society for Music Information
  Retrieval (ISMIR), 2003, Baltimore,USA.} {Proc. 4th international society for
  music information retrieval (ismir), 2003, baltimore,usa.}
\PrintBackRefs{\CurrentBib}

\bibitem [\protect \citeauthoryear {%
Gogineni%
, Kuriakose%
\BCBL {}\ \BBA {} Murthy%
}{%
Gogineni%
\ \protect \BOthers {.}}{%
{\protect \APACyear {2018}}%
}]{%
gogineni2018mridangam}
\APACinsertmetastar {%
gogineni2018mridangam}%
\begin{APACrefauthors}%
Gogineni, K.%
, Kuriakose, J.%
\BCBL {}\ \BBA {} Murthy, H\BPBI A.%
\end{APACrefauthors}%
\unskip\
\newblock
\APACrefYearMonthDay{2018}{}{}.
\newblock
{\BBOQ}\APACrefatitle {Mridangam Artist Identification from Taniavartanam
  Audio} {Mridangam artist identification from taniavartanam audio}.{\BBCQ}
\newblock
\BIn{} \APACrefbtitle {Twenty Fourth National Conference on Communications
  (NCC) 2018} {Twenty fourth national conference on communications (ncc) 2018}\
  (\BPGS\ 1--6).
\PrintBackRefs{\CurrentBib}

\bibitem [\protect \citeauthoryear {%
Gottlieb%
}{%
Gottlieb%
}{%
{\protect \APACyear {1993}}%
}]{%
gottlieb1993solo}
\APACinsertmetastar {%
gottlieb1993solo}%
\begin{APACrefauthors}%
Gottlieb, R\BPBI S.%
\end{APACrefauthors}%
\unskip\
\newblock
\APACrefYear{1993}.
\newblock
\APACrefbtitle {Solo tabla drumming of North India: Its repertoire, styles, and
  performance practices} {Solo tabla drumming of north india: Its repertoire,
  styles, and performance practices}.
\newblock
\APACaddressPublisher{}{Motilal Banarsidass Publishers}.
\PrintBackRefs{\CurrentBib}

\bibitem [\protect \citeauthoryear {%
Gowriprasad%
\ \BBA {} Murty%
}{%
Gowriprasad%
\ \BBA {} Murty%
}{%
{\protect \APACyear {2020}}%
}]{%
gowriprasad2020onset}
\APACinsertmetastar {%
gowriprasad2020onset}%
\begin{APACrefauthors}%
Gowriprasad, R.%
\BCBT {}\ \BBA {} Murty, K\BPBI S\BPBI R.%
\end{APACrefauthors}%
\unskip\
\newblock
\APACrefYearMonthDay{2020}{}{}.
\newblock
{\BBOQ}\APACrefatitle {Onset detection of Tabla Strokes using LP Analysis}
  {Onset detection of tabla strokes using lp analysis}.{\BBCQ}
\newblock
\BIn{} \APACrefbtitle {International Conference on Signal Processing and
  Communications (SPCOM)} {International conference on signal processing and
  communications (spcom)}\ (\BPGS\ 1--5).
\PrintBackRefs{\CurrentBib}

\bibitem [\protect \citeauthoryear {%
Gowriprasad%
, Venkatesh%
, Murthy%
, Aravind%
\BCBL {}\ \BBA {} Murty%
}{%
Gowriprasad%
\ \protect \BOthers {.}}{%
{\protect \APACyear {2021}}%
}]{%
gowriprasad_r_gharana_ISMIR2021}
\APACinsertmetastar {%
gowriprasad_r_gharana_ISMIR2021}%
\begin{APACrefauthors}%
Gowriprasad, R.%
, Venkatesh, V.%
, Murthy, H\BPBI A.%
, Aravind, R.%
\BCBL {}\ \BBA {} Murty, K\BPBI S\BPBI R.%
\end{APACrefauthors}%
\unskip\
\newblock
\APACrefYearMonthDay{2021}{{\APACmonth{11}}}{}.
\newblock
{\BBOQ}\APACrefatitle {{Tabla Gharana Recognition from Audio music recordings
  of Tabla Solo performances}} {{Tabla Gharana Recognition from Audio music
  recordings of Tabla Solo performances}}.{\BBCQ}
\newblock
\BIn{} \APACrefbtitle {{Proceedings of the 22nd International Society for Music
  Information Retrieval Conference}} {{Proceedings of the 22nd International
  Society for Music Information Retrieval Conference}}\ (\BPG~547-554).
\newblock
\APACaddressPublisher{Online}{ISMIR}.
\newblock
\begin{APACrefURL} \url{https://doi.org/10.5281/zenodo.5624631}
  \end{APACrefURL}
\newblock
\begin{APACrefDOI} \doi{10.5281/zenodo.5624631} \end{APACrefDOI}
\PrintBackRefs{\CurrentBib}

\bibitem [\protect \citeauthoryear {%
Gowriprasad%
, Venkatesh%
\BCBL {}\ \BBA {} Murty~K%
}{%
Gowriprasad%
\ \protect \BOthers {.}}{%
{\protect \APACyear {2022}}%
}]{%
gowriprasad_r_gharana_NCC2022}
\APACinsertmetastar {%
gowriprasad_r_gharana_NCC2022}%
\begin{APACrefauthors}%
Gowriprasad, R.%
, Venkatesh, V.%
\BCBL {}\ \BBA {} Murty~K, S\BPBI R.%
\end{APACrefauthors}%
\unskip\
\newblock
\APACrefYearMonthDay{2022}{}{}.
\newblock
{\BBOQ}\APACrefatitle {Tabla gharana Recognition from Tabla Solo Recordings}
  {Tabla gharana recognition from tabla solo recordings}.{\BBCQ}
\newblock
\BIn{} \APACrefbtitle {2022 National Conference on Communications (NCC)} {2022
  national conference on communications (ncc)}\ (\BPG~59-63).
\newblock
\begin{APACrefDOI} \doi{10.1109/NCC55593.2022.9806767} \end{APACrefDOI}
\PrintBackRefs{\CurrentBib}

\bibitem [\protect \citeauthoryear {%
Grosche%
, M{\"u}ller%
\BCBL {}\ \BBA {} Kurth%
}{%
Grosche%
\ \protect \BOthers {.}}{%
{\protect \APACyear {2010}}%
}]{%
grosche2010cyclic}
\APACinsertmetastar {%
grosche2010cyclic}%
\begin{APACrefauthors}%
Grosche, P.%
, M{\"u}ller, M.%
\BCBL {}\ \BBA {} Kurth, F.%
\end{APACrefauthors}%
\unskip\
\newblock
\APACrefYearMonthDay{2010}{}{}.
\newblock
{\BBOQ}\APACrefatitle {Cyclic tempogram—A mid-level tempo representation for
  musicsignals} {Cyclic tempogram—a mid-level tempo representation for
  musicsignals}.{\BBCQ}
\newblock
\BIn{} \APACrefbtitle {2010 IEEE International Conference on Acoustics, Speech
  and Signal Processing} {2010 ieee international conference on acoustics,
  speech and signal processing}\ (\BPGS\ 5522--5525).
\PrintBackRefs{\CurrentBib}

\bibitem [\protect \citeauthoryear {%
Gupta%
, Srinivasamurthy%
, Kumar%
, Murthy%
\BCBL {}\ \BBA {} Serra%
}{%
Gupta%
\ \protect \BOthers {.}}{%
{\protect \APACyear {2015}}%
}]{%
gupta2015discovery}
\APACinsertmetastar {%
gupta2015discovery}%
\begin{APACrefauthors}%
Gupta, S.%
, Srinivasamurthy, A.%
, Kumar, M.%
, Murthy, H\BPBI A.%
\BCBL {}\ \BBA {} Serra, X.%
\end{APACrefauthors}%
\unskip\
\newblock
\APACrefYearMonthDay{2015}{}{}.
\newblock
{\BBOQ}\APACrefatitle {Discovery of syllabic percussion patterns in tabla solo
  recordings} {Discovery of syllabic percussion patterns in tabla solo
  recordings}.{\BBCQ}
\newblock
\BIn{} \APACrefbtitle {Proc. 16th International Society for Music Information
  Retrieval (ISMIR); 2015 Oct 26-30; M{\'a}laga, Spain.[M{\'a}laga]. p.
  385-391.} {Proc. 16th international society for music information retrieval
  (ismir); 2015 oct 26-30; m{\'a}laga, spain.[m{\'a}laga]. p. 385-391.}
\PrintBackRefs{\CurrentBib}

\bibitem [\protect \citeauthoryear {%
Ioffe%
\ \BBA {} Szegedy%
}{%
Ioffe%
\ \BBA {} Szegedy%
}{%
{\protect \APACyear {2015}}%
}]{%
ioffe2015batch}
\APACinsertmetastar {%
ioffe2015batch}%
\begin{APACrefauthors}%
Ioffe, S.%
\BCBT {}\ \BBA {} Szegedy, C.%
\end{APACrefauthors}%
\unskip\
\newblock
\APACrefYearMonthDay{2015}{}{}.
\newblock
{\BBOQ}\APACrefatitle {Batch normalization: Accelerating deep network training
  by reducing internal covariate shift} {Batch normalization: Accelerating deep
  network training by reducing internal covariate shift}.{\BBCQ}
\newblock
\BIn{} \APACrefbtitle {International conference on machine learning}
  {International conference on machine learning}\ (\BPGS\ 448--456).
\PrintBackRefs{\CurrentBib}

\bibitem [\protect \citeauthoryear {%
Jensen%
}{%
Jensen%
}{%
{\protect \APACyear {2006}}%
}]{%
jensen2006multiple}
\APACinsertmetastar {%
jensen2006multiple}%
\begin{APACrefauthors}%
Jensen, K.%
\end{APACrefauthors}%
\unskip\
\newblock
\APACrefYearMonthDay{2006}{}{}.
\newblock
{\BBOQ}\APACrefatitle {Multiple scale music segmentation using rhythm, timbre,
  and harmony} {Multiple scale music segmentation using rhythm, timbre, and
  harmony}.{\BBCQ}
\newblock
\APACjournalVolNumPages{EURASIP Journal on Advances in Signal
  Processing}{2007}{}{1--11}.
\PrintBackRefs{\CurrentBib}

\bibitem [\protect \citeauthoryear {%
Kingma%
\ \BBA {} Ba%
}{%
Kingma%
\ \BBA {} Ba%
}{%
{\protect \APACyear {2014}}%
}]{%
kingma2014adam}
\APACinsertmetastar {%
kingma2014adam}%
\begin{APACrefauthors}%
Kingma, D\BPBI P.%
\BCBT {}\ \BBA {} Ba, J.%
\end{APACrefauthors}%
\unskip\
\newblock
\APACrefYearMonthDay{2014}{}{}.
\newblock
{\BBOQ}\APACrefatitle {Adam: A method for stochastic optimization} {Adam: A
  method for stochastic optimization}.{\BBCQ}
\newblock
\APACjournalVolNumPages{arXiv preprint arXiv:1412.6980}{}{}{}.
\PrintBackRefs{\CurrentBib}

\bibitem [\protect \citeauthoryear {%
Klapuri%
, Virtanen%
, Eronen%
\BCBL {}\ \BBA {} Sepp{\"a}nen%
}{%
Klapuri%
\ \protect \BOthers {.}}{%
{\protect \APACyear {2001}}%
}]{%
klapuri2001automatic}
\APACinsertmetastar {%
klapuri2001automatic}%
\begin{APACrefauthors}%
Klapuri, A.%
, Virtanen, T.%
, Eronen, A.%
\BCBL {}\ \BBA {} Sepp{\"a}nen, J.%
\end{APACrefauthors}%
\unskip\
\newblock
\APACrefYearMonthDay{2001}{}{}.
\newblock
{\BBOQ}\APACrefatitle {Automatic transcription of musical recordings}
  {Automatic transcription of musical recordings}.{\BBCQ}
\newblock
\BIn{} \APACrefbtitle {Consistent \& Reliable Acoustic Cues Workshop, CRAC-01,
  Aalborg, Denmark.} {Consistent \& reliable acoustic cues workshop, crac-01,
  aalborg, denmark.}
\PrintBackRefs{\CurrentBib}

\bibitem [\protect \citeauthoryear {%
Kuriakose%
, Kumar%
, Sarala%
, Murthy%
\BCBL {}\ \BBA {} Sivaraman%
}{%
Kuriakose%
\ \protect \BOthers {.}}{%
{\protect \APACyear {2015}}%
}]{%
kuriakose2015akshara}
\APACinsertmetastar {%
kuriakose2015akshara}%
\begin{APACrefauthors}%
Kuriakose, J.%
, Kumar, J\BPBI C.%
, Sarala, P.%
, Murthy, H\BPBI A.%
\BCBL {}\ \BBA {} Sivaraman, U\BPBI K.%
\end{APACrefauthors}%
\unskip\
\newblock
\APACrefYearMonthDay{2015}{}{}.
\newblock
{\BBOQ}\APACrefatitle {Akshara transcription of mrudangam strokes in carnatic
  music} {Akshara transcription of mrudangam strokes in carnatic music}.{\BBCQ}
\newblock
\BIn{} \APACrefbtitle {Twenty First National Conference on Communications (NCC)
  2015.} {Twenty first national conference on communications (ncc) 2015.}
\PrintBackRefs{\CurrentBib}

\bibitem [\protect \citeauthoryear {%
MA%
, TP%
\BCBL {}\ \BBA {} Rao%
}{%
MA%
\ \protect \BOthers {.}}{%
{\protect \APACyear {2020}}%
}]{%
ma2020structural}
\APACinsertmetastar {%
ma2020structural}%
\begin{APACrefauthors}%
MA, R.%
, TP, V.%
\BCBL {}\ \BBA {} Rao, P.%
\end{APACrefauthors}%
\unskip\
\newblock
\APACrefYearMonthDay{2020}{}{}.
\newblock
{\BBOQ}\APACrefatitle {Structural segmentation of dhrupad vocal bandish audio
  based on tempo} {Structural segmentation of dhrupad vocal bandish audio based
  on tempo}.{\BBCQ}
\newblock
\BIn{} \APACrefbtitle {Proc. of int. soc. for music information retrieval
  conf.(ismir).(Montreal, Canada).} {Proc. of int. soc. for music information
  retrieval conf.(ismir).(montreal, canada).}
\PrintBackRefs{\CurrentBib}

\bibitem [\protect \citeauthoryear {%
Nair%
\ \BBA {} Hinton%
}{%
Nair%
\ \BBA {} Hinton%
}{%
{\protect \APACyear {2010}}%
}]{%
nair2010rectified}
\APACinsertmetastar {%
nair2010rectified}%
\begin{APACrefauthors}%
Nair, V.%
\BCBT {}\ \BBA {} Hinton, G\BPBI E.%
\end{APACrefauthors}%
\unskip\
\newblock
\APACrefYearMonthDay{2010}{}{}.
\newblock
{\BBOQ}\APACrefatitle {Rectified linear units improve restricted boltzmann
  machines} {Rectified linear units improve restricted boltzmann
  machines}.{\BBCQ}
\newblock
\BIn{} \APACrefbtitle {Icml.} {Icml.}
\PrintBackRefs{\CurrentBib}

\bibitem [\protect \citeauthoryear {%
of Primary%
\ \BBA {} Education%
}{%
of Primary%
\ \BBA {} Education%
}{%
{\protect \APACyear {2002}}%
}]{%
Tabla_Vidwat}
\APACinsertmetastar {%
Tabla_Vidwat}%
\begin{APACrefauthors}%
of Primary, D.%
\BCBT {}\ \BBA {} Education, S.%
\end{APACrefauthors}%
\unskip\
\newblock
\APACrefYear{2002}.
\newblock
\APACrefbtitle {Hindustani Tabla, Vidwat Purva Hanta ( Per- Proficiency Grade)}
  {Hindustani tabla, vidwat purva hanta ( per- proficiency grade)}.
\newblock
\APACaddressPublisher{}{Government of Karnataka}.
\PrintBackRefs{\CurrentBib}

\bibitem [\protect \citeauthoryear {%
Oord%
, Li%
\BCBL {}\ \BBA {} Vinyals%
}{%
Oord%
\ \protect \BOthers {.}}{%
{\protect \APACyear {2018}}%
}]{%
oord2018representation}
\APACinsertmetastar {%
oord2018representation}%
\begin{APACrefauthors}%
Oord, A\BPBI v\BPBI d.%
, Li, Y.%
\BCBL {}\ \BBA {} Vinyals, O.%
\end{APACrefauthors}%
\unskip\
\newblock
\APACrefYearMonthDay{2018}{}{}.
\newblock
{\BBOQ}\APACrefatitle {Representation learning with contrastive predictive
  coding} {Representation learning with contrastive predictive coding}.{\BBCQ}
\newblock
\APACjournalVolNumPages{arXiv preprint arXiv:1807.03748}{}{}{}.
\PrintBackRefs{\CurrentBib}

\bibitem [\protect \citeauthoryear {%
Padi%
\ \BBA {} Murthy%
}{%
Padi%
\ \BBA {} Murthy%
}{%
{\protect \APACyear {2018}}%
}]{%
padi2018segmentation}
\APACinsertmetastar {%
padi2018segmentation}%
\begin{APACrefauthors}%
Padi, S.%
\BCBT {}\ \BBA {} Murthy, H\BPBI A.%
\end{APACrefauthors}%
\unskip\
\newblock
\APACrefYearMonthDay{2018}{}{}.
\newblock
{\BBOQ}\APACrefatitle {Segmentation of continuous audio recordings of Carnatic
  music concerts into items for archival} {Segmentation of continuous audio
  recordings of carnatic music concerts into items for archival}.{\BBCQ}
\newblock
\APACjournalVolNumPages{S{\=a}dhan{\=a}}{43}{10}{1--20}.
\PrintBackRefs{\CurrentBib}

\bibitem [\protect \citeauthoryear {%
Paulus%
, M{\"u}ller%
\BCBL {}\ \BBA {} Klapuri%
}{%
Paulus%
\ \protect \BOthers {.}}{%
{\protect \APACyear {2010}}%
}]{%
paulus2010state}
\APACinsertmetastar {%
paulus2010state}%
\begin{APACrefauthors}%
Paulus, J.%
, M{\"u}ller, M.%
\BCBL {}\ \BBA {} Klapuri, A.%
\end{APACrefauthors}%
\unskip\
\newblock
\APACrefYearMonthDay{2010}{}{}.
\newblock
{\BBOQ}\APACrefatitle {State of the art report: Audio-based music structure
  analysis.} {State of the art report: Audio-based music structure
  analysis.}{\BBCQ}
\newblock
\BIn{} \APACrefbtitle {Proc. 11th International Society for Music Information
  Retrieval (ISMIR)} {Proc. 11th international society for music information
  retrieval (ismir)}\ (\BPG~625–636).
\PrintBackRefs{\CurrentBib}

\bibitem [\protect \citeauthoryear {%
Peeters%
}{%
Peeters%
}{%
{\protect \APACyear {2003}}%
}]{%
peeters2003deriving}
\APACinsertmetastar {%
peeters2003deriving}%
\begin{APACrefauthors}%
Peeters, G.%
\end{APACrefauthors}%
\unskip\
\newblock
\APACrefYearMonthDay{2003}{}{}.
\newblock
{\BBOQ}\APACrefatitle {Deriving musical structures from signal analysis for
  music audio summary generation:“sequence” and “state” approach}
  {Deriving musical structures from signal analysis for music audio summary
  generation:“sequence” and “state” approach}.{\BBCQ}
\newblock
\BIn{} \APACrefbtitle {International Symposium on Computer Music Modeling and
  Retrieval} {International symposium on computer music modeling and
  retrieval}\ (\BPGS\ 143--166).
\PrintBackRefs{\CurrentBib}

\bibitem [\protect \citeauthoryear {%
Pradhan%
}{%
Pradhan%
}{%
{\protect \APACyear {2011}}%
}]{%
pradhan2011tabla}
\APACinsertmetastar {%
pradhan2011tabla}%
\begin{APACrefauthors}%
Pradhan, A.%
\end{APACrefauthors}%
\unskip\
\newblock
\APACrefYear{2011}.
\newblock
\APACrefbtitle {Tabla: A Performer's Perspective} {Tabla: A performer's
  perspective}.
\newblock
\APACaddressPublisher{}{BookBaby}.
\PrintBackRefs{\CurrentBib}

\bibitem [\protect \citeauthoryear {%
PV%
, Sankaran%
\BCBL {}\ \BBA {} Murthy%
}{%
PV%
\ \protect \BOthers {.}}{%
{\protect \APACyear {2016}}%
}]{%
pv2016segmentation}
\APACinsertmetastar {%
pv2016segmentation}%
\begin{APACrefauthors}%
PV, K\BPBI S.%
, Sankaran, S.%
\BCBL {}\ \BBA {} Murthy, H.%
\end{APACrefauthors}%
\unskip\
\newblock
\APACrefYearMonthDay{2016}{}{}.
\newblock
{\BBOQ}\APACrefatitle {Segmentation of Carnatic music items using KL2, GMM and
  CFB energy feature} {Segmentation of carnatic music items using kl2, gmm and
  cfb energy feature}.{\BBCQ}
\newblock
\BIn{} \APACrefbtitle {2016 Twenty Second National Conference on Communication
  (NCC)} {2016 twenty second national conference on communication (ncc)}\
  (\BPGS\ 1--5).
\PrintBackRefs{\CurrentBib}

\bibitem [\protect \citeauthoryear {%
Ranjani%
\ \BBA {} Sreenivas%
}{%
Ranjani%
\ \BBA {} Sreenivas%
}{%
{\protect \APACyear {2013}}%
}]{%
ranjani2013hierarchical}
\APACinsertmetastar {%
ranjani2013hierarchical}%
\begin{APACrefauthors}%
Ranjani, H.%
\BCBT {}\ \BBA {} Sreenivas, T.%
\end{APACrefauthors}%
\unskip\
\newblock
\APACrefYearMonthDay{2013}{}{}.
\newblock
{\BBOQ}\APACrefatitle {Hierarchical classification of carnatic music forms}
  {Hierarchical classification of carnatic music forms}.{\BBCQ}
\newblock

\PrintBackRefs{\CurrentBib}

\bibitem [\protect \citeauthoryear {%
Rao%
, Vinutha%
\BCBL {}\ \BBA {} Rohit%
}{%
Rao%
\ \protect \BOthers {.}}{%
{\protect \APACyear {2020}}%
}]{%
rao2020structural}
\APACinsertmetastar {%
rao2020structural}%
\begin{APACrefauthors}%
Rao, P.%
, Vinutha, T\BPBI P.%
\BCBL {}\ \BBA {} Rohit, M\BPBI A.%
\end{APACrefauthors}%
\unskip\
\newblock
\APACrefYearMonthDay{2020}{}{}.
\newblock
{\BBOQ}\APACrefatitle {Structural segmentation of alap in Dhrupad vocal
  concerts} {Structural segmentation of alap in dhrupad vocal concerts}.{\BBCQ}
\newblock
\APACjournalVolNumPages{Transactions of the International Society for Music
  Information Retrieval}{3}{1}{}.
\PrintBackRefs{\CurrentBib}

\bibitem [\protect \citeauthoryear {%
Rohit%
\ \BBA {} Rao%
}{%
Rohit%
\ \BBA {} Rao%
}{%
{\protect \APACyear {2018}}%
}]{%
rohitacoustic}
\APACinsertmetastar {%
rohitacoustic}%
\begin{APACrefauthors}%
Rohit, M.%
\BCBT {}\ \BBA {} Rao, P.%
\end{APACrefauthors}%
\unskip\
\newblock
\APACrefYearMonthDay{2018}{}{}.
\newblock
{\BBOQ}\APACrefatitle {Acoustic-Prosodic Features of Tabla Bol Recitation and
  Correspondence with the Tabla Imitation.} {Acoustic-prosodic features of
  tabla bol recitation and correspondence with the tabla imitation.}{\BBCQ}
\newblock
\BIn{} \APACrefbtitle {Interspeech} {Interspeech}\ (\BPGS\ 1229--1233).
\PrintBackRefs{\CurrentBib}

\bibitem [\protect \citeauthoryear {%
Rohit%
\ \BBA {} Rao%
}{%
Rohit%
\ \BBA {} Rao%
}{%
{\protect \APACyear {2020}}%
}]{%
rohit2020structure}
\APACinsertmetastar {%
rohit2020structure}%
\begin{APACrefauthors}%
Rohit, M.%
\BCBT {}\ \BBA {} Rao, P.%
\end{APACrefauthors}%
\unskip\
\newblock
\APACrefYearMonthDay{2020}{}{}.
\newblock
{\BBOQ}\APACrefatitle {Structure and automatic segmentation of Dhrupad vocal
  bandish audio} {Structure and automatic segmentation of dhrupad vocal bandish
  audio}.{\BBCQ}
\newblock
\APACjournalVolNumPages{Unpublished technical report}{}{}{}.
\PrintBackRefs{\CurrentBib}

\bibitem [\protect \citeauthoryear {%
Samudravijaya%
, Shah%
\BCBL {}\ \BBA {} Pandya%
}{%
Samudravijaya%
\ \protect \BOthers {.}}{%
{\protect \APACyear {2004}}%
}]{%
samudravijaya2004computer}
\APACinsertmetastar {%
samudravijaya2004computer}%
\begin{APACrefauthors}%
Samudravijaya, K.%
, Shah, S.%
\BCBL {}\ \BBA {} Pandya, P.%
\end{APACrefauthors}%
\unskip\
\newblock
\APACrefYearMonthDay{2004}{}{}.
\newblock
\APACrefbtitle {Computer recognition of tabla bols} {Computer recognition of
  tabla bols}\ \APACbVolEdTR{}{\BTR{}}.
\newblock
\APACaddressInstitution{}{Technical report, Tata Institute of Fundamental
  Research}.
\PrintBackRefs{\CurrentBib}

\bibitem [\protect \citeauthoryear {%
Sankaran%
, Krishnaraj~Sekhar%
\BCBL {}\ \BBA {} Hema%
}{%
Sankaran%
\ \protect \BOthers {.}}{%
{\protect \APACyear {2015}}%
}]{%
sankaran2015automatic}
\APACinsertmetastar {%
sankaran2015automatic}%
\begin{APACrefauthors}%
Sankaran, S.%
, Krishnaraj~Sekhar, P.%
\BCBL {}\ \BBA {} Hema, A\BPBI M.%
\end{APACrefauthors}%
\unskip\
\newblock
\APACrefYearMonthDay{2015}{}{}.
\newblock
{\BBOQ}\APACrefatitle {Automatic segmentation of composition in carnatic music
  using time-frequency cfcc templates} {Automatic segmentation of composition
  in carnatic music using time-frequency cfcc templates}.{\BBCQ}
\newblock
\BIn{} \APACrefbtitle {Proceedings of 11th International Symposium on Computer
  Music Multidisciplinary Research.} {Proceedings of 11th international
  symposium on computer music multidisciplinary research.}
\PrintBackRefs{\CurrentBib}

\bibitem [\protect \citeauthoryear {%
Saxena%
}{%
Saxena%
}{%
{\protect \APACyear {2006}}%
}]{%
saxena2006art}
\APACinsertmetastar {%
saxena2006art}%
\begin{APACrefauthors}%
Saxena, S\BPBI K.%
\end{APACrefauthors}%
\unskip\
\newblock
\APACrefYear{2006}.
\newblock
\APACrefbtitle {The Art of Tabl{\=a} Rhythm: Essentials, Tradition, and
  Creativity} {The art of tabl{\=a} rhythm: Essentials, tradition, and
  creativity}\ (\BNUM~8).
\newblock
\APACaddressPublisher{}{Sangeet Natak Akademi}.
\PrintBackRefs{\CurrentBib}

\bibitem [\protect \citeauthoryear {%
Smith%
, Burgoyne%
, Fujinaga%
, De~Roure%
\BCBL {}\ \BBA {} Downie%
}{%
Smith%
\ \protect \BOthers {.}}{%
{\protect \APACyear {2011}}%
}]{%
smith2011design}
\APACinsertmetastar {%
smith2011design}%
\begin{APACrefauthors}%
Smith, J\BPBI B\BPBI L.%
, Burgoyne, J\BPBI A.%
, Fujinaga, I.%
, De~Roure, D.%
\BCBL {}\ \BBA {} Downie, J\BPBI S.%
\end{APACrefauthors}%
\unskip\
\newblock
\APACrefYearMonthDay{2011}{}{}.
\newblock
{\BBOQ}\APACrefatitle {Design and creation of a large-scale database of
  structural annotations.} {Design and creation of a large-scale database of
  structural annotations.}{\BBCQ}
\newblock
\BIn{} \APACrefbtitle {ISMIR} {Ismir}\ (\BVOL~11, \BPGS\ 555--560).
\PrintBackRefs{\CurrentBib}

\bibitem [\protect \citeauthoryear {%
Sobel%
}{%
Sobel%
}{%
{\protect \APACyear {2014}}%
}]{%
Sobel_filter}
\APACinsertmetastar {%
Sobel_filter}%
\begin{APACrefauthors}%
Sobel, I.%
\end{APACrefauthors}%
\unskip\
\newblock
\APACrefYearMonthDay{2014}{02}{}.
\newblock
{\BBOQ}\APACrefatitle {An Isotropic 3x3 Image Gradient Operator} {An isotropic
  3x3 image gradient operator}.{\BBCQ}
\newblock
\APACjournalVolNumPages{Presentation at Stanford A.I. Project 1968}{}{}{}.
\PrintBackRefs{\CurrentBib}

\bibitem [\protect \citeauthoryear {%
Srinivasamurthy%
, Holzapfel%
\BCBL {}\ \BBA {} Serra%
}{%
Srinivasamurthy%
\ \protect \BOthers {.}}{%
{\protect \APACyear {2014}}%
}]{%
srinivasamurthy2014search}
\APACinsertmetastar {%
srinivasamurthy2014search}%
\begin{APACrefauthors}%
Srinivasamurthy, A.%
, Holzapfel, A.%
\BCBL {}\ \BBA {} Serra, X.%
\end{APACrefauthors}%
\unskip\
\newblock
\APACrefYearMonthDay{2014}{}{}.
\newblock
{\BBOQ}\APACrefatitle {In search of automatic rhythm analysis methods for
  turkish and indian art music} {In search of automatic rhythm analysis methods
  for turkish and indian art music}.{\BBCQ}
\newblock
\APACjournalVolNumPages{Journal of New Music Research}{43}{1}{94--114}.
\PrintBackRefs{\CurrentBib}

\bibitem [\protect \citeauthoryear {%
Thoshkahna%
, M{\"u}ller%
, Kulkarni%
\BCBL {}\ \BBA {} Jiang%
}{%
Thoshkahna%
\ \protect \BOthers {.}}{%
{\protect \APACyear {2015}}%
}]{%
thoshkahna2015novel}
\APACinsertmetastar {%
thoshkahna2015novel}%
\begin{APACrefauthors}%
Thoshkahna, B.%
, M{\"u}ller, M.%
, Kulkarni, V.%
\BCBL {}\ \BBA {} Jiang, N.%
\end{APACrefauthors}%
\unskip\
\newblock
\APACrefYearMonthDay{2015}{}{}.
\newblock
{\BBOQ}\APACrefatitle {Novel audio features for capturing tempo salience in
  music recordings} {Novel audio features for capturing tempo salience in music
  recordings}.{\BBCQ}
\newblock
\BIn{} \APACrefbtitle {2015 IEEE International Conference on Acoustics, Speech
  and Signal Processing (ICASSP)} {2015 ieee international conference on
  acoustics, speech and signal processing (icassp)}\ (\BPGS\ 181--185).
\PrintBackRefs{\CurrentBib}

\bibitem [\protect \citeauthoryear {%
Turnbull%
, Lanckriet%
, Pampalk%
\BCBL {}\ \BBA {} Goto%
}{%
Turnbull%
\ \protect \BOthers {.}}{%
{\protect \APACyear {2007}}%
}]{%
turnbull2007supervised}
\APACinsertmetastar {%
turnbull2007supervised}%
\begin{APACrefauthors}%
Turnbull, D.%
, Lanckriet, G\BPBI R.%
, Pampalk, E.%
\BCBL {}\ \BBA {} Goto, M.%
\end{APACrefauthors}%
\unskip\
\newblock
\APACrefYearMonthDay{2007}{}{}.
\newblock
{\BBOQ}\APACrefatitle {A Supervised Approach for Detecting Boundaries in Music
  Using Difference Features and Boosting.} {A supervised approach for detecting
  boundaries in music using difference features and boosting.}{\BBCQ}
\newblock
\BIn{} \APACrefbtitle {Proc. 8th International Society for Music Information
  Retrieval (ISMIR)} {Proc. 8th international society for music information
  retrieval (ismir)}\ (\BPGS\ 51--54).
\PrintBackRefs{\CurrentBib}

\bibitem [\protect \citeauthoryear {%
Ullrich%
, Schl{\"u}ter%
\BCBL {}\ \BBA {} Grill%
}{%
Ullrich%
\ \protect \BOthers {.}}{%
{\protect \APACyear {2014}}%
}]{%
ullrich2014boundary}
\APACinsertmetastar {%
ullrich2014boundary}%
\begin{APACrefauthors}%
Ullrich, K.%
, Schl{\"u}ter, J.%
\BCBL {}\ \BBA {} Grill, T.%
\end{APACrefauthors}%
\unskip\
\newblock
\APACrefYearMonthDay{2014}{}{}.
\newblock
{\BBOQ}\APACrefatitle {Boundary Detection in Music Structure Analysis using
  Convolutional Neural Networks.} {Boundary detection in music structure
  analysis using convolutional neural networks.}{\BBCQ}
\newblock
\BIn{} \APACrefbtitle {ISMIR} {Ismir}\ (\BPGS\ 417--422).
\PrintBackRefs{\CurrentBib}

\bibitem [\protect \citeauthoryear {%
Verma%
, Vinutha%
, Pandit%
\BCBL {}\ \BBA {} Rao%
}{%
Verma%
\ \protect \BOthers {.}}{%
{\protect \APACyear {2015}}%
}]{%
verma2015structural}
\APACinsertmetastar {%
verma2015structural}%
\begin{APACrefauthors}%
Verma, P.%
, Vinutha, T.%
, Pandit, P.%
\BCBL {}\ \BBA {} Rao, P.%
\end{APACrefauthors}%
\unskip\
\newblock
\APACrefYearMonthDay{2015}{}{}.
\newblock
{\BBOQ}\APACrefatitle {Structural segmentation of Hindustani concert audio with
  posterior features} {Structural segmentation of hindustani concert audio with
  posterior features}.{\BBCQ}
\newblock
\BIn{} \APACrefbtitle {2015 IEEE International Conference on Acoustics, Speech
  and Signal Processing (ICASSP)} {2015 ieee international conference on
  acoustics, speech and signal processing (icassp)}\ (\BPGS\ 136--140).
\PrintBackRefs{\CurrentBib}

\bibitem [\protect \citeauthoryear {%
Vidwans%
, Ganguli%
\BCBL {}\ \BBA {} Rao%
}{%
Vidwans%
\ \protect \BOthers {.}}{%
{\protect \APACyear {2012}}%
}]{%
vidwans2012classification}
\APACinsertmetastar {%
vidwans2012classification}%
\begin{APACrefauthors}%
Vidwans, A.%
, Ganguli, K\BPBI K.%
\BCBL {}\ \BBA {} Rao, P.%
\end{APACrefauthors}%
\unskip\
\newblock
\APACrefYearMonthDay{2012}{}{}.
\newblock
{\BBOQ}\APACrefatitle {Classification of indian classical vocal styles from
  melodic contours} {Classification of indian classical vocal styles from
  melodic contours}.{\BBCQ}
\newblock
\BIn{} \APACrefbtitle {Serra X, Rao P, Murthy H, Bozkurt B, editors.
  Proceedings of the 2nd CompMusic Workshop; 2012 Jul 12-13; Istanbul, Turkey.
  Barcelona: Universitat Pompeu Fabra; 2012. p. 139-146.} {Serra x, rao p,
  murthy h, bozkurt b, editors. proceedings of the 2nd compmusic workshop; 2012
  jul 12-13; istanbul, turkey. barcelona: Universitat pompeu fabra; 2012. p.
  139-146.}
\PrintBackRefs{\CurrentBib}

\bibitem [\protect \citeauthoryear {%
Vidwans%
, Verma%
\BCBL {}\ \BBA {} Rao%
}{%
Vidwans%
\ \protect \BOthers {.}}{%
{\protect \APACyear {2020}}%
}]{%
vidwans2020classifying}
\APACinsertmetastar {%
vidwans2020classifying}%
\begin{APACrefauthors}%
Vidwans, A.%
, Verma, P.%
\BCBL {}\ \BBA {} Rao, P.%
\end{APACrefauthors}%
\unskip\
\newblock
\APACrefYearMonthDay{2020}{}{}.
\newblock
{\BBOQ}\APACrefatitle {Classifying Cultural Music using Melodic Features}
  {Classifying cultural music using melodic features}.{\BBCQ}
\newblock
\BIn{} \APACrefbtitle {2020 International Conference on Signal Processing and
  Communications (SPCOM)} {2020 international conference on signal processing
  and communications (spcom)}\ (\BPGS\ 1--5).
\PrintBackRefs{\CurrentBib}

\bibitem [\protect \citeauthoryear {%
Vinutha%
, Sankagiri%
, Ganguli%
\BCBL {}\ \BBA {} Rao%
}{%
Vinutha%
, Sankagiri%
, Ganguli%
\BCBL {}\ \BBA {} Rao%
}{%
{\protect \APACyear {2016}}%
}]{%
vinutha2016structural}
\APACinsertmetastar {%
vinutha2016structural}%
\begin{APACrefauthors}%
Vinutha, T.%
, Sankagiri, S.%
, Ganguli, K\BPBI K.%
\BCBL {}\ \BBA {} Rao, P.%
\end{APACrefauthors}%
\unskip\
\newblock
\APACrefYearMonthDay{2016}{}{}.
\newblock
{\BBOQ}\APACrefatitle {Structural Segmentation and Visualization of Sitar and
  Sarod Concert Audio.} {Structural segmentation and visualization of sitar and
  sarod concert audio.}{\BBCQ}
\newblock
\BIn{} \APACrefbtitle {ISMIR} {Ismir}\ (\BPGS\ 232--238).
\PrintBackRefs{\CurrentBib}

\bibitem [\protect \citeauthoryear {%
Vinutha%
, Sankagiri%
\BCBL {}\ \BBA {} Rao%
}{%
Vinutha%
, Sankagiri%
\BCBL {}\ \BBA {} Rao%
}{%
{\protect \APACyear {2016}}%
}]{%
vinutha2016reliable}
\APACinsertmetastar {%
vinutha2016reliable}%
\begin{APACrefauthors}%
Vinutha, T.%
, Sankagiri, S.%
\BCBL {}\ \BBA {} Rao, P.%
\end{APACrefauthors}%
\unskip\
\newblock
\APACrefYearMonthDay{2016}{}{}.
\newblock
{\BBOQ}\APACrefatitle {Reliable tempo detection for structural segmentation in
  sarod concerts} {Reliable tempo detection for structural segmentation in
  sarod concerts}.{\BBCQ}
\newblock
\BIn{} \APACrefbtitle {2016 Twenty Second National Conference on Communication
  (NCC)} {2016 twenty second national conference on communication (ncc)}\
  (\BPGS\ 1--6).
\PrintBackRefs{\CurrentBib}

\bibitem [\protect \citeauthoryear {%
Yong%
, Choi%
\BCBL {}\ \BBA {} Nam%
}{%
Yong%
\ \protect \BOthers {.}}{%
{\protect \APACyear {2020}}%
}]{%
yong2020pytsmod}
\APACinsertmetastar {%
yong2020pytsmod}%
\begin{APACrefauthors}%
Yong, S.%
, Choi, S.%
\BCBL {}\ \BBA {} Nam, J.%
\end{APACrefauthors}%
\unskip\
\newblock
\APACrefYearMonthDay{2020}{}{}.
\newblock
{\BBOQ}\APACrefatitle {PyTSMod: A python implementation of time-scale
  modification algorithm} {Pytsmod: A python implementation of time-scale
  modification algorithm}.{\BBCQ}
\newblock
\BIn{} \APACrefbtitle {Extended Abstracts for the Late-Breaking Demo Session of
  the 21th International Society for Music InformationRetrieval Conference
  (ISMIR), 2020.} {Extended abstracts for the late-breaking demo session of the
  21th international society for music informationretrieval conference (ismir),
  2020.}
\PrintBackRefs{\CurrentBib}

\end{thebibliography}

\newpage



\end{document}